\documentclass[]{emulateapj}
\usepackage{graphicx, amsmath, amsthm, amssymb}

\def\ba{\begin{eqnarray}}
\def\ea{\end{eqnarray}}
\shorttitle{Coplanar High-eccentricity Migration}
\shortauthors{Petrovich}

\begin{document}

\title{Hot Jupiters from 
coplanar high-eccentricity migration.}
\author{Cristobal Petrovich\altaffilmark{1}}
\altaffiltext{1}{Department of Astrophysical Sciences, Princeton University, Ivy Lane, Princeton, NJ 08544, USA; cpetrovi@princeton.edu}

\begin{abstract}
We study the possibility that hot Jupiters  are formed 
through the secular gravitational interactions between two planets 
in eccentric orbits 
with relatively low mutual inclinations ($\lesssim20^\circ$)
and friction due to tides raised on the planet by the host star.
We term this migration mechanism Coplanar High-eccentricity
Migration because, like disk migration, it allows for migration
to occur on the same plane in which the planets formed.
Coplanar High-eccentricity Migration can operate
from the following typical initial configurations:
(i) inner planet in a circular orbit and the outer planet with 
an eccentricity $\gtrsim0.67$ for
$m_{\rm in}/m_{\rm out}(a_{\rm in}/a_{\rm out})^{1/2}\lesssim0.3$;
(ii) two eccentric ($\gtrsim0.5$) orbits
for $m_{\rm in}/m_{\rm out}(a_{\rm in}/a_{\rm out})^{1/2}\lesssim0.16$.
A population synthesis study 
of hierarchical systems of two giant planets
using the observed eccentricity distribution
of giant planets shows that  Coplanar High-eccentricity Migration 
produces hot Jupiters
 with low stellar obliquities ($\lesssim30^\circ$),
with a semi-major axis distribution that matches the
observations, and at a rate that can account for
their observed occurrence.
A different mechanism is needed to create large obliquity 
hot Jupiters, either a different migration channel or a mechanism
that tilts the star or the proto-planetary disk. 
Coplanar High-eccentricity Migration  
predicts that hot Jupiters
should have distant ($a\gtrsim5$ AU) and 
massive (most likely $\sim1-3$ more massive than the hot Jupiter)
companions with relatively low mutual inclinations 
($\lesssim 20^\circ$) and moderately high 
eccentricities ($e\sim0.2-0.5$).
\end{abstract}                                     
\keywords{planetary systems --
 planets and satellites: dynamical evolution and stability --
  planets and satellites: formation }

\section{Introduction}
\label{sec:intro}

At least $\sim10-15\%$ of Sun-like stars harbor a 
Jovian-mass planet,
while only $\sim0.5-1\%$ harbor a 
so-called  hot Jupiter (HJ) with semi-major axis $<0.1$ AU
 \citep{marcy05,gould06,mayor11,wright12,howard12}.
Both radial velocity (RV) and transit surveys show that the HJs
are piled up at a semi-major axis of
$\sim0.04-0.05$ AU (e.g., \citealt{hellier12}), 
and some of the HJs have significant eccentricities
($\sim 10\%$ have $e>0.2$) and stellar obliquities: 
$\sim 30\%$ of HJs have projected spin-orbit misalignment angle 
$\lambda>30^\circ$ as determined
by Rossiter-MacLaughlin measurements\footnote{Taken from The 
Exoplanet Orbit Database and a sample with 
$M\sin i>0.1M_J$\citep{wright11} }.

Hot Jupiters could  not have formed at their 
current locations because of the high gas temperature 
and low disk mass at these small radii
\citep{bod10}.
Instead, they must have formed beyond a few AU
and then have migrated inwards, probably by angular momentum 
exchange with the protoplanetary disk \citep{GT80,W97} 
or by high-eccentricity migration (e.g., \citealt{FR96,WM03}), 
in which the migrating planet
attains very high eccentricities and tidal dissipation circularizes
the orbit.
Within the latter migration scenario, several different
mechanisms to excite the eccentricity to high values
 have been proposed:
the Kozai-Lidov mechanism in stellar binaries
\citep{WM03,FT07,naoz12,petro15}, 
planet-planet scattering \citep{FR96,NIB08,NI11,BN12},
and secular interactions between planets 
\citep{WL11,naoz11}.
Although all of the migration mechanisms above 
can form hot Jupiters, which  is the
dominant channel (if any) remains an open question.

In this paper, we study the possibility that hot Jupiters are formed
by the secular interaction of two planets in initially eccentric orbits
in a hierarchical configuration
($a_{\rm in}\ll [1-e_{\rm out}]a_{\rm out}$) 
with relatively low mutual inclinations ($\lesssim20^\circ$).
We term this migration mechanism 
``Coplanar High-eccentricity Migration" (CHEM) to
differentiate it from previously proposed high-eccentricity 
migration channels in
which the eccentricity and inclination excitation generally
go hand-in-hand (e.g., \citealt{NIB08,FT07,naoz11}).

Dynamically unstable multiple-planet systems 
generally relax into a long-term stable 
configuration with two planets in eccentric
and hierarchical orbits  \citep{WM96,LI97,JT08,CFMS2008}.
The eccentricity distribution of these systems 
can reproduce the wide eccentricity distribution 
(median of $\simeq0.23$) observed in the RV sample 
\citep{FR08,JT08,CFMS2008}.
Planet-planet scattering does not only excite the planetary 
eccentricities, but it does also excite the planetary inclinations. 
However, in a significant fraction of the reported
 scattering experiments the planets end up in orbits with 
 $e\gtrsim0.5$ and mutual inclinations $\lesssim20^\circ$ 
($0.35$ radians), for which CHEM
can  operate \citep{timpe13,JT08,CFMS2008}.
 In particular,  \citet{timpe13} show that the mutual inclination
of two surviving planets after planet-planet scattering in an initial 
three-planet system follows an exponential distribution with mean 
$\simeq3.4^\circ-5.7^\circ$ ($\simeq0.06-0.1$).

Various systems of two hierarchical planets
on eccentric orbits are known to date.
\citet{KR14} show that four known RV planets in multi-planet
systems exhibit large amplitude
secular eccentricity oscillations (inner planet reaches
a maximum eccentricity $\sim0.6-0.8$) and a stability analysis
suggests that their (unknown) mutual inclinations are 
not too high so the inner planet avoids plunging into star.
Similarly, \citet{dawson14} show that Kepler-419 is a hierarchical 
system ($a_{\rm in}=0.37$ AU, $a_{\rm out}=1.68$ AU)
where the inner and outer planets have eccentricities
of  $\simeq0.83$ and $\simeq0.184$, respectively, 
while their mutual inclination is $9^{+8}_{-6}$ degrees.


The secular interaction between two planets in a hierarchical
and coplanar configuration has been previously studied by
several authors (e.g., \citealt{malhotra02,LP03,MM04,LH05,MFB06,MM09}).
In particular,  \citet{LP03} and \citet{MM04} show that
that the planetary orbits 
can engage in libration of 
$\varpi\equiv\varpi_{\rm in}-\varpi_{\rm out}$
around either $0^\circ$ or $180^\circ$, where $\varpi_{\rm in}$ 
and $\varpi_{\rm out}$ are the
longitudes of pericenter of the inner and outer bodies. 
This libration can cause large amplitude
eccentricity oscillations of either planet and, most
important for this work, in some cases
the inner planet might reach eccentricities large
enough for friction due to tides raised on the planet by the host star
to become important.

Similar to the previous work by \citet{LP03} and 
\citet{MM04}, 
\citet{li14} recently studied the secular evolution of two hierarchical, 
nearly coplanar, and eccentric bodies, but in the test particle limit
($m_{\rm in}/m_{\rm out}\ll1$).
These authors confirmed that in this limit, the inner planet can 
reach unit eccentricity, derived a simple analytical condition 
for this to happen (see Eq. [\ref{eq:li14}]), and showed
that the orbit can flip its angular momentum vector
to produce a coplanar retrograde planet.

Here,  we extend these works by studying the conditions on 
the masses and the orbital elements 
in hierarchical and nearly coplanar planetary systems
required to drive the eccentricities close to unity, and also
by including the effects from general relativistic precession
and tides that can limit the eccentricity growth.
 
\section{Analytic results}
\label{sec:analytic}

In this section we use a time-averaged Hamiltonian 
of  two hierarchical and nearly coplanar orbits 
expanded in series of the semi-major axis ratio 
to describe their secular evolution
and assess which orbital elements and planetary masses 
allow for $e_{\rm in}\to 1$.

As discussed by \citet{LP03} the coplanar problem has one
degree of freedom: three variables 
($e_{\rm in}$, $e_{\rm out}$,
and $\varpi\equiv\varpi_{\rm in}-\varpi_{\rm out}$) and two 
conserved quantities 
(energy and total orbital angular momentum,
 Equations [\ref{eq:potential}] and [\ref{eq:am}]).

Hereafter, we shall use the notation from \citet{petro15} in 
which the variables are the eccentricity vectors ${\bf e}_{\rm in}$
and ${\bf e}_{\rm out}$, and the orbital angular momentum vectors
${\bf h}_{\rm in}$ and ${\bf h}_{\rm out}$ all defined in the 
Jacobi's reference frame\footnote{The description of the reference
frame in the Appendix A of \citet{petro15} has a typo 
and should say:
``We define the inner orbit relative to bodies 1 and 2, while the outer 
orbit is defined relative to bodies 3 
and the center of mass of bodies 1 and 2".
In this paper, body 1 is the host star and bodies 2
 and 3 are the inner and outer planets.}.
We denote the masses of the central star and inner 
(outer) planets as $m_1$ and $m_{\rm in}$
($m_{\rm out}$), respectively.

The double time averaged interaction potential 
in the octupole approximation (expansion
up to $a_{\rm{in}}^3/a_{\rm{out}}^4$)
is ${\phi}_{\rm oct}=\tilde{\phi}_{\rm{oct}}\phi_0$, where 
 in the planetary limit ($ m_{\rm in},m_{\rm out}\ll m_1$) 
we have
\citep{petro15}:
\ba
\tilde{\phi}_{\rm{oct}}&=&\frac{\phi_{\rm oct}}{\phi_0}=
\frac{e_{\rm{in}}^2+2/3}{2(1-e_{\rm{out}}^2)^{3/2}}-
\frac{5\alpha}{16}\frac{3e_{\rm{in}}^2+4}{(1-e_{\rm{out}}^2)^{5/2}} 
\bf{e}_{\rm{in}}\cdot \bf{e}_{\rm{out}},\nonumber\\
\label{eq:potential}
\ea
with
\ba
\phi_0=\frac{3Gm_{\rm in}m_{\rm out}a_{\rm in}^2}{4a_{\rm out}^3}, 
\label{eq:phi_0}
\ea
$\alpha=a_{\rm in}/a_{\rm out}$, and 
$\bf{\hat{e}}_{\rm{in}}\cdot \bf{\hat{e}}_{\rm{out}}=\cos\varpi$. 

This potential is accurate to first order
in the mutual inclination $i_{\rm tot}$
with $\cos i_{\rm tot}=\hat{{\bf h}}_{\rm in}\cdot\hat{{\bf h}}_{\rm out}$
and it has proven to be very accurate for
 $\alpha\lesssim0.1$ in the planetary limit \citep{LP03}.
Note that this interaction potential has positive energy---the 
opposite sign as the standard definition of the
interaction Hamiltonian.

Similarly, we define the ratio between the total orbital angular 
momentum and the total orbital angular momentum 
that would obtain if the orbits were circular as
\ba
\mathcal{J}=\frac{\mu\alpha^{1/2}(1-e_{\rm{in}}^2)^{1/2}
+(1-e_{\rm{out}}^2)^{1/2}}{\mu\alpha^{1/2}+1},
\label{eq:am}
\ea
where $\mu $ is the planetary mass ratio
$\mu=m_{\rm in}/m_{\rm out}$. 

The quantity $\mathcal{J}$ is a constant of motion in the 
secular approximation and in the absence 
of extra forces other than the gravitational interactions between 
the planets and the star.
This result immediately implies that for a given $\mathcal{J}$ 
we have 
\ba 
e_{\rm in}\leq \sqrt{1-\left[\frac{(1+\mu\alpha^{1/2})\mathcal{J}-1}
{\mu\alpha^{1/2}}\right]^2}
\label{eq:e_lim}
\ea
if $(1+\mu\alpha^{1/2})\mathcal{J}\geq1$,
while $e_{\rm in}$ can reach unity if
$(1+\mu\alpha^{1/2})\mathcal{J}\leq1$
(set $e_{\rm in}=1$ and $e_{\rm out}\geq0$ in
Eq. [\ref{eq:am}]).

\begin{figure*}
   \centering
  \includegraphics[width=18cm]{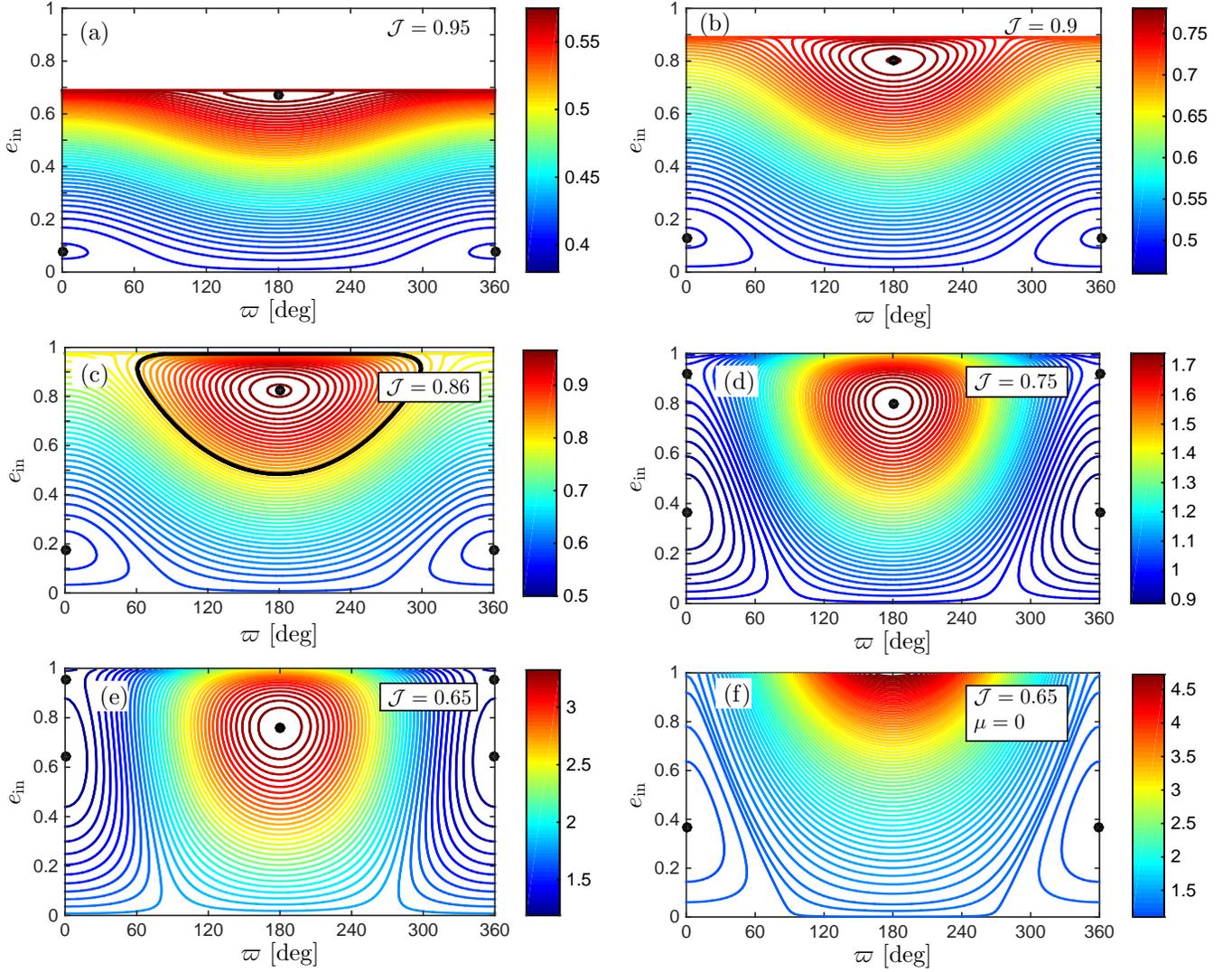} 
  \caption{Level curves of the dimensionless potential 
  $\tilde{\phi}_{\rm oct}$ (Eq. \ref{eq:potential}) for different values
of $\mathcal{J}$ (Eq. \ref{eq:am}) as indicated in each 
  panel. 
  In panels a-e we fix $\alpha\equiv a_{\rm in}/a_{\rm out}=1/8$ 
  and $\mu\equiv m_{\rm in}/m_{\rm out}=0.606$ as in the example 
in Figure \ref{fig:example_sec}, and in panel c we show the
trajectory with $\mathcal{J}=0.86$ and 
$\tilde{\phi}_{\rm oct}=0.83$ that corresponds to the initial
condition of that example (black thick line).
In panel f we show the test particle limit $\mu=0$ with
$\alpha=1/8$ and $\mathcal{J}=0.65$.
The black dots indicate the fixed points of 
$de_{\rm in}/dt=d\varpi/dt=0$ from
Equations (\ref{eq:e_dot}) and (\ref{eq:omega_dot}).
  }
\label{fig:contour}
\end{figure*}  

\subsection{Phase-space trajectories}

In Figure \ref{fig:contour} we show level curves of 
the dimensionless potential 
  $\tilde{\phi}_{\rm oct}$ in Equation (\ref{eq:potential}) 
  for different values of the dimensionless
  total orbital angular momentum 
  $\mathcal{J}$ in Equation (\ref{eq:am}).
From panels a to e, we fix the planetary mass ratio to $\mu=0.606$ 
and the semi-major axis ratio to $\alpha=1/8$, similar to our example in
  Figure \ref{fig:example_sec}.
  For these parameters $e_{\rm in}$ can reach unity
  if $\mathcal{J}\leq0.824$.
 In panel f we show the test particle limit $\mu=0$
 with $\alpha=1/8$ and $\mathcal{J}=0.65$.
  In all panels, we indicate the fixed points 
 $de_{\rm in}/dt=d\varpi/dt=0$ from
Equations (\ref{eq:e_dot}) and (\ref{eq:omega_dot}) as
black circles. 

From panel a, we observe that for $\mathcal{J}=0.95$ 
the phase-space trajectories are restricted 
to $e_{\rm in}\leq0.697$ as required from
Equation  (\ref{eq:e_lim}).
Most trajectories correspond to circulation of the 
relative apsidal angle
$\varpi$ and the minimum 
eccentricities happen at  $\varpi=180^\circ$ 
(anti-parallel eccentricity vectors).
The eccentricity variation between $\varpi=180^\circ$ and 
$\varpi=0$ (or $\varpi=360^\circ$) is 
at most $\sim0.15$.
There are two fixed points of the equations of motions
(solutions to $de_{\rm in}/dt=d\varpi/dt=0$ in
Equations [\ref{eq:e_dot}] and [\ref{eq:omega_dot}]):
one at $\varpi=0$ and $e_{\rm in}=0.0782$
with low energy ($\tilde{\phi}_{\rm oct}=0.405$),
and another at 
$\varpi=180^\circ$ and $e_{\rm in}=0.681$
with high energy ($\tilde{\phi}_{\rm oct}=0.591$).
Close to these fixed points the trajectories correspond to 
librations of $\varpi$ and the eccentricity,
as previously identified by \citet{LP03}.

By decreasing the total orbital angular momentum from
$\mathcal{J}=0.95$  to $\mathcal{J}=0.9$ (panel b), the
fixed point at $\varpi=180^\circ$
moves from $e_{\rm in}=0.681$ to 
$e_{\rm in}=0.81$, while that at $\varpi=0$ moves
from $e_{\rm in}=0.0782$ to $e_{\rm in}=0.128$.
The libration region around these two fixed points occupies
a larger volume in $e_{\rm in}-\varpi$ space relative to 
that when $\mathcal{J}=0.95$. 
The angular momentum constraint in Equation  
(\ref{eq:e_lim}) limits the eccentricity to $e_{\rm in}<0.901$.

In panel c, we decrease the angular momentum even further 
to $\mathcal{J}=0.86$ (panel c), which allows for a
maximum eccentricity $e_{\rm in}=0.978$ (Eq. [\ref{eq:e_lim}]).
The parameters in this panel are chosen to coincide
with the initial conditions from our example 
in Figure \ref{fig:example_sec} in which the inner planet 
undergoes migration.
That phase-space trajectory in this example is indicated
by the black thick line and it corresponds to 
large amplitude eccentricity librations in the
range $e_{\rm in}\simeq0.5-0.98$ and 
$\varpi$ in $\simeq 60^\circ-300^\circ$.
We note that for the energy levels close to our example 
($\tilde{\phi}_{\rm oct}\sim0.8$)
there are trajectories that could lead to eccentricities
close to unity from either circulation or libration
of $\varpi$.

In panel d we set $\mathcal{J}=0.75$ and 
observe that most trajectories with 
$\tilde{\phi}_{\rm oct}\lesssim1.6$ pass through 
$e_{\rm in}\simeq1$.
Even if one starts from a circular orbit and 
$\varpi\sim100^\circ-260^\circ$ the eccentricity
of the inner planet always attains very high values.
Also, we observe that there are two fixed points at 
$\varpi=0$: one at $e_{\rm in}=0.359$ and the 
other at  $e_{\rm in}=0.922$.
The former corresponds to a stable fix point around which 
$\varpi$ librates with possibly large amplitude
eccentricity oscillations, while the latter is unstable 
(a saddle point) and it only appears when 
$\mathcal{J}<0.845$.

In panel e we set $\mathcal{J}=0.65$.
By decreasing $\mathcal{J}$ from 0.75 to 0.65 we observe
that the fixed points at $\varpi=0$ move to higher values
and that the trajectories starting from circular orbits 
can reach unity eccentricities for all values of 
$\varpi$.

 In panel f we show the test particle limit $\mu=0$ for 
 $\mathcal{J}=0.65$ (or $e_{\rm out}$=0.76 from Eq. [\ref{eq:am}]) 
and observe that there is only one fixed point 
 at $\varpi=0$ and $e_{\rm in}=0.365$.
Consistently,  from Equation (\ref{eq:omega_dot}) one can 
easily show that for all values of $\mathcal{J}$
there is no physical solution for 
 $d\varpi/dt=0$  when $\varpi=180^\circ$.
 Similarly, there is only one physical solution
 when $\varpi=0$, which is given by
$ e_{\rm in}=\left(1-\sqrt{1-4\beta^2}\right)/(3\beta)$
with $\beta=15\alpha\sqrt{1-\mathcal{J}^2}/(8\mathcal{J})$.
This result implies that in the test particle approximation
there can be only libration of $\varpi$ and $e_{\rm in}$
around one fixed point at $\varpi=0$.

In summary, the secular phase-space trajectories of two hierarchical
and coplanar orbits include circulation of $\varpi$
 and also libration of $\varpi$ around $0$ 
 and $180^\circ$. Both the circulating and the librating trajectories
around $180^\circ$ can lead to very high values of
 $e_{\rm in}$.
In the test particle approximation the libration of 
$\varpi$ around $180^\circ$ is not present.

\subsection{Available phase-space 
for migration}
\label{sec:mig_space}

Hereafter, we use the subscripts $i$ and $f$ to denote 
the initial and final states.

In the test particle approximation, $e_{\rm out}$ is constant
and Equation (\ref{eq:potential}) implies that
$e_{{\rm in},f}\to1$ 
 only if 
\ba	
\alpha\frac{e_{\rm out}}{1-e_{\rm out}^2}=
\frac{8}{5}\frac{1-e_{{\rm in},i}^2}{7\cos \varpi_f
-e_{{\rm in},i}(4+3e_{{\rm in},i}^2)
\cos\varpi_i},
\label{eq:li14}
\ea
which translates into the condition of
\citet{li14}  (Equation 14 therein) since 
$\cos\varpi_f\leq1$.

In what follows, we do not assume that the inner
planet is a test particle.

\begin{figure}
   \centering
  \includegraphics[width=8.7cm]{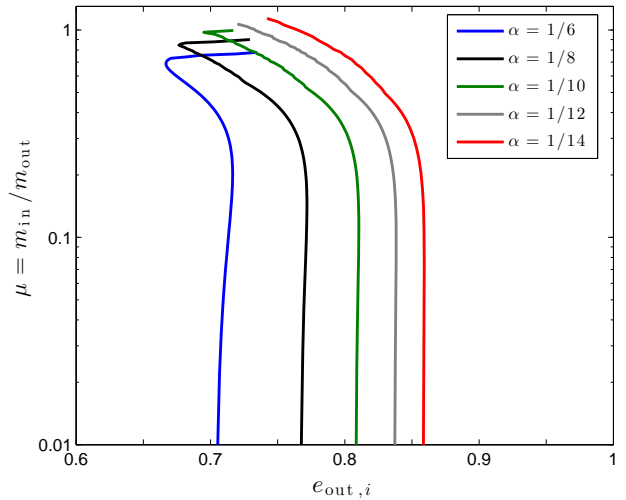} 
  \caption{Solutions to Equation (\ref{eq:cond_full}) 
 as a function of the initial eccentricity of the outer planet 
 $e_{{\rm out},i}$
and the planetary mass ratio $\mu=m_{\rm in}/m_{\rm out}$
(see text).
The color label indicates different values of the semi-major
axis ratio $\alpha=\{1/6,1/8,1/10,1/12,1/14\}$.
The solutions correspond to the minimum eccentricity 
of the outer planet that is required for the inner planet to
increase its eccentricity from 0 to 1 for a given value
of $\mu$.}
\label{fig:e_max_mass}
\end{figure}  

\subsubsection{Initial circular orbit}
\label{sec:circular}
Let us  start by assuming that in the initial state $e_{{\rm in},i}=0$ 
and it reaches a final state with $e_{{\rm in},f}\to1$. Thus, the 
energy conservation in Equation  (\ref{eq:potential})
implies
\ba
\frac{1}{3(1-e_{{\rm out},i}^2)^{3/2}}&=&
\frac{5}{6(1-e_{{\rm out},f}^2)^{3/2}}  \nonumber \\
&-&\frac{35\alpha}{16}\frac{e_{{\rm out},f}}{(1-e_{{\rm out},f}^2)^{5/2}} 
\cos\varpi_f,
\label{eq:cond_full_0}
\ea
and by using 
$\mu\alpha^{1/2}+(1-e_{{\rm out},i}^2)^{1/2}=(1-e_{{\rm out},f}^2)^{1/2}$
we get a condition for $e_{{\rm out},i}$, $\mu$, $\varpi_f$,
and $\alpha$, as:
\ba
\frac{1}{3(1-e_{{\rm out},i}^2)^{3/2}}=
\frac{5}{6 \left[\mu\alpha^{1/2}+(1-e_{{\rm out},i}^2)^{1/2}\right]^3}\nonumber\\
-\frac{35\alpha}{16}\frac{\left\{1- \left[\mu\alpha^{1/2}+
(1-e_{{\rm out},i}^2)^{1/2}\right]^2\right\}^{1/2}}
{ \left[\mu\alpha^{1/2}+(1-e_{{\rm out},i}^2)^{1/2}\right]^5}\cos\varpi_f,
\label{eq:cond_full}
\ea
where we note that the minimum (maximum) value of 
$e_{{\rm out},i}$ required to 
solve this equation is given by $\varpi_f=0$ ($\varpi_f=\pi$).
However, it can happen that a phase-space trajectory connecting
$e_{{\rm in},i}=0$ with $e_{{\rm in},f}\to1$ and $\varpi_f=0$ 
might not exist.
Thus, in order to find the minimum outer eccentricity to reach 
$e_{{\rm in},f}\to1$ we numerically find
the minimum value of $\varpi_f$ (if any) that connects 
$e_{\rm in}=0$ with $e_{\rm in}=1$, while satisfying Equation 
(\ref{eq:cond_full}). 
As an example, from panel d in Figure \ref{fig:contour} the path
that connects $e_{\rm in}=0$ with $e_{\rm in}=1$ has
$\varpi_f\simeq50^\circ$.

We proceed as follows. For each combination of $\mu$ and $\alpha$
we solve the Equation (\ref{eq:cond_full}) starting with $\varpi_f=0$
and check if the phase-space trajectory is continuous. If the trajectory is continuous,
then we have determined the minimum eccentricity of the
outer planet to reach $e_{{\rm in},f}\to1$. 
If trajectory is not continuous, we increase $\varpi_f$
and repeat the procedure until we find a continuous path (if any) with 
$\varpi_f=0-180^\circ$.  

In Figure \ref{fig:e_max_mass} we show our results  
for the minimum initial eccentricity of the outer
planet $e_{{\rm out}, i,{\rm min}}$
to excite $e_{\rm in}$ from 0 to 1 as a function of the planetary
mass ratio $\mu$ and for different values of the semi-major
axis ratio $\alpha$.
We observe that for a fix value of $\alpha$, 
$e_{{\rm out}, i,{\rm min}}$ reaches its lowest value of $\sim0.67-0.75$
for $\mu\sim0.7-1$, while it increases almost monotonically
for lower values of $\mu$.
Similarly, $e_{{\rm out}, i,{\rm min}}$ increases as $\alpha$
decreases. 
We describe these observations below.

In the test particle approximation 
 the trajectories connecting 
$e_{{\rm in},i}=0$ with $e_{{\rm in},f}=1$ are all continuous
(the fixed point at $\varpi=0$ and high $e_{\rm in}$
disappears), implying that the minimum $\varpi_f$ is 0 
(see panel f in Figure \ref{fig:contour}).
Thus, by setting $\varpi_f=0$ in Equation (\ref{eq:cond_full_0})
the minimum eccentricity $e_{{\rm out}, i,{\rm min}}$  in the
test particle approximation (constant $e_{\rm out}$) 
is given by 
\ba	
e_{{\rm out},i, {\rm min}}=\frac{\sqrt{1+\gamma^2}-1}{\gamma},
\label{eq:li14_v2}
\ea
where $\gamma=16/(35\alpha)$.


When the inner and outer masses are comparable, the
the eccentricity of the outer planet can change.
We can calculate $e_{{\rm out}, i,{\rm min}}$ from
the limiting case in which $e_{{\rm out}, f}=0$:
the inner orbit transfers the maximum angular momentum 
possible to the outer orbit.
Thus, by setting  $e_{{\rm out}, f}=0$ in 
Equation (\ref{eq:cond_full_0}) we obtain
\ba
e_{{\rm out}, i,{\rm min}}=\sqrt{1-\left(2/5\right)^{2/3}}
=0.676,
\ea
which roughly coincides with the lowest
values of  $e_{{\rm out}, i,{\rm min}}$ in Figure 
 \ref{fig:e_max_mass} for $\mu\sim0.7-1$
 and $\alpha\leq1/10$.

The values of $\mu$ and $\alpha$
at which $e_{{\rm out}, i,{\rm min}}$ is lowest
can be estimated by setting 
$e_{{\rm out}, i,{\rm min}}=0.676$ and 
$e_{{\rm out}, f}=0$ in the angular 
momentum conservation condition,
which results in
$\mu \alpha^{1/2}=1-\left( 2/5\right)^{1/3}=0.263$.
This value is only an estimate and we numerically find that  
$0.3$ approximates better than $0.263$ 
the position of the minimum $e_{{\rm out}, i,{\rm min}}$
in Figure \ref{fig:e_max_mass}.
Thus, we conclude from this analysis
that  the parameters required
to excite the eccentricity 
of the inner planet from zero to unity 
with the lowest eccentricities
of the outer planet should satisfy:
 \ba
 \mu \alpha^{1/2}\equiv\frac{m_{\rm in}}{m_{\rm out}}
 \left(\frac{a_{\rm in}}{a_{\rm out}}\right)^{1/2}\simeq0.3.
 \label{eq:mu_alpha}
\ea
For $\mu \alpha^{1/2}\gtrsim0.3$ there are no solutions to
Equation (\ref{eq:cond_full_0}), while for 
$\mu \alpha^{1/2}\lesssim0.3$ the required eccentricities 
increase with decreasing $\mu \alpha^{1/2}$ 
until they reach the test particle limit ($\mu\ll1$), which is
given by Equation (\ref{eq:li14_v2}).


We note that given the high values of $e_{{\rm out},i}$ required 
to reach $e_{\rm in}\to1$, starting from 
a circular orbit might cause the system to become dynamically 
unstable.
According to the stability boundary of hierarchical triple systems
from \citet{MA01} (Eq. [\ref{eq:stability}]), a planetary system 
($ m_{\rm in},m_{\rm out}\ll m_1$) with an outer eccentricity 
of $e_{\rm out }=0.676$ is stable for $\alpha<1/13.3$.
Thus, if the systems with $\alpha>1/13.3$ were indeed 
unstable the available phase-space
for migration starting from an inner circular orbit would
be strongly limited.
However, in a recent work \citet{petro15b} shows that
most systems with $m_{\rm in}\sim m_{\rm out}\sim 1M_J$, $e_{\rm in}\sim0$
and $e_{\rm out}\simeq0.7$ are long-term
stable for $\alpha\lesssim1/8$. We adopt this less
conservative stability limit for our discussion in \S\ref{sec:outer}.

In summary, the eccentricity excitation of the inner planet
from a circular to a radial orbit is possible only if the outer 
body starts from an eccentric orbit with 
$e_{\rm out}\gtrsim0.67$
and the mass and semi-major axis ratios
satisfy $\mu \alpha^{1/2}\lesssim0.3$.
As $\mu \alpha^{1/2}$ departs from 0.3 the required 
eccentricities of the outer planet increase, implying that
the eccentricity excitation is most efficient for planets 
of comparable masses with 
$\mu=m_{\rm in}/m_{\rm out}\sim0.7-1$.

\subsubsection{Initial eccentric orbit}
\label{sec:ini_ecc}

We now relax the requirement that the inner planet
is initially in a circular orbit. 
Thus, we use the conservation of energy in Equation 
(\ref{eq:potential}) and only fix $\varpi_i=\pi$ as
\ba 
\tilde{\phi}_{\rm{oct}}(e_{{\rm in},i},e_{{\rm out},i},\varpi_{i}=\pi)
=\tilde{\phi}_{\rm{oct}}(e_{{\rm in},f}=1,e_{{\rm out},f},\varpi_{f}),
\nonumber\\
\label{eq:cond_full_2}
\ea
which can be numerically solved along with the
angular momentum conservation condition
in Equation (\ref{eq:am}) for different values of
of $e_{{\rm in},i}$ and $\varpi_f$.

\begin{figure}
   \centering
  \includegraphics[width=8.5cm]{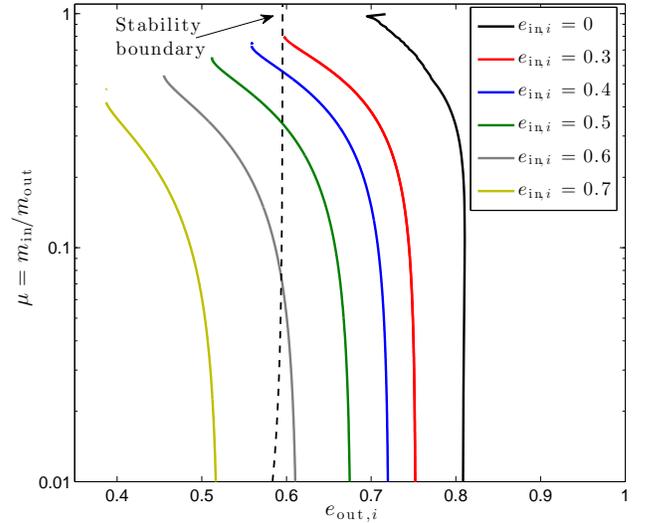} 
  \caption{Solutions to Equation (\ref{eq:cond_full_2}) 
  as a function of the initial 
 eccentricity of the outer planet $e_{{\rm out},i}$
and the planetary mass ratio $\mu=m_{\rm in}/m_{\rm out}$,
for a fixed semi-major axis ratio $\alpha=1/10$.
The color labels indicate different values of the initial
eccentricity of the inner planet $e_{{\rm in},i}$.
The solutions correspond to the minimum eccentricity 
of the outer planet that is required for the inner planet to
increase its eccentricity from  $e_{{\rm in},i}$ to 1
for a given value of $\mu$.
The stability boundary for hierarchical triple
systems from Equation (\ref{eq:stability})
is indicated as a dashed black line (stable configurations
are left to the line).}
\label{fig:e_max_mass_e}
\end{figure}  

\begin{figure*}
   \centering
  \includegraphics[width=17cm]{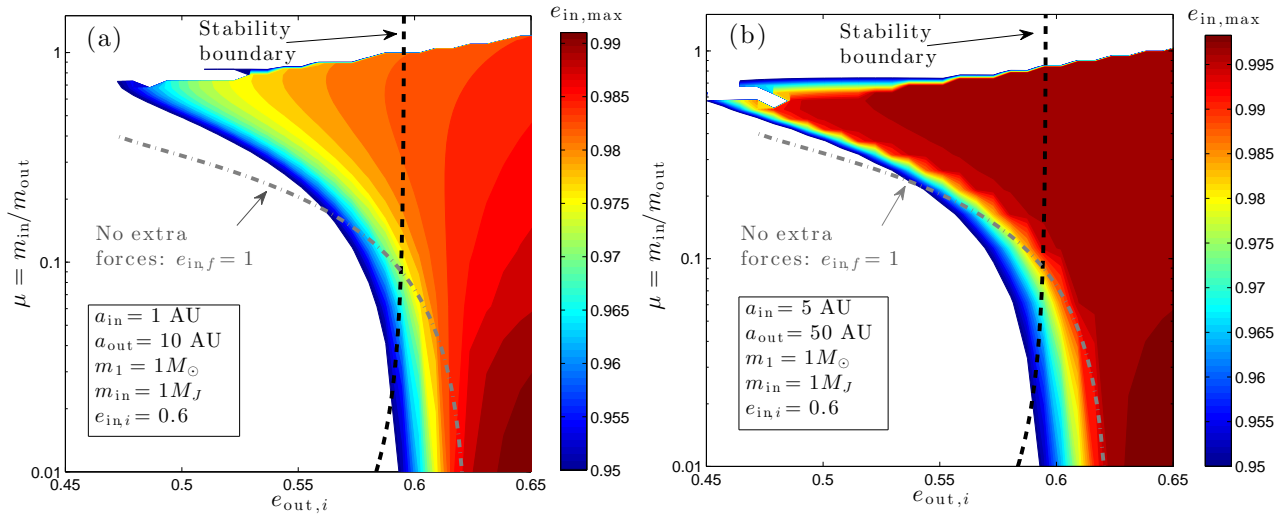} 
  \caption{Contour levels of the maximum eccentricity 
of the inner planet as a function of $e_{{\rm out},i}$
and the mass ratio $\mu=m_{\rm in}/m_{\rm out}$ obtained
from Equations (\ref{eq:am}) and (\ref{eq:cond_full_3}),
which include the extra forces from 
general relativity and the tidal quadrupole.
We fix the initial eccentricity of the inner planet to $e_{{\rm in},i}=0.6$
and the masses of the star and the inner planet to $m_1=1M_\odot$
and $m_{\rm in}=1M_J$, respectively.
{\it Panel a:} the semi-major axes of the inner and outer planets
are $a_{\rm in}=1$ AU and $a_{\rm out}=10$ AU. 
{\it Panel b:} the semi-major axes of the inner and outer planets
are $a_{\rm in}=5$ AU and $a_{\rm out}=50$ AU. 
The stability boundary from Equation (\ref{eq:stability}) is
shown as the dashed black line.
The minimum initial eccentricity of the outer planet required to
reach $e_{\rm in}=1$ when no extra forces are included
(Eq. [\ref{eq:cond_full_2}]) is shown as the dot-dashed gray line.
}
\label{fig:emax_phi}
\end{figure*}  

In Figure \ref{fig:e_max_mass_e} we show the roots 
of Equation (\ref{eq:cond_full_2}) 
numerically minimizing the outer eccentricity $e_{{\rm out},i}$ over
$\varpi_f$ restricted to continuous phase-space 
trajectories (see \S\ref{sec:circular})
for different values
of the initial eccentricity of the inner planet.
For a given mass ratio $\mu$ each curve indicates
the minimum eccentricity of the outer planet that
is required to excite the eccentricity of the inner planet
from $e_{{\rm in},i}$ to $1$. 

Not surprisingly, we observe from this figure that by 
starting from higher initial 
eccentricities of the inner planet we require smaller 
eccentricities of the outer planet to reach 
$e_{{\rm in},f}\simeq1$, as expected.
Also, as we increase $e_{{\rm in},i}$ the maximum
mass ratio $\mu$ at which the eccentricity excitation can happen
is lower and the minimum values of the outer eccentricity 
$e_{{\rm out},i}$ are reached when $\mu\sim0.3-1$.

The analysis can be further simplified by assuming that 
initially the eccentricities of inner and outer planets are equal:
$e_{{\rm in},i}=e_{{\rm out},i}$.
This is an arbitrary assumption that we use to derive analytical 
expressions.
Similar to the previous section, we can determine
the initial minimum eccentricity (of the inner and outer planets) 
$e_{{\rm min}}$ required to reach $e_{{\rm in},f}\to1$ by observing that the
maximum angular momentum transfer from the inner 
to the outer orbit occurs when $e_{{\rm out}, f}\to0$.
Replacing these limits in Equation (\ref{eq:cond_full_2}),
we get
\ba
\frac{e_{\rm{min}}^2+2/3}{2(1-e_{\rm{min}}^2)^{3/2}}+
\frac{5\alpha}{16}\frac{(3e_{\rm{min}}^2+4)e_{\rm{min}}^2}
{(1-e_{\rm{min}}^2)^{5/2}} 
=\frac{5}{6};
\label{eq:cond_ecc}
\ea
in the limit $\alpha\to0$ the zero of this equation
is $e_{\rm{min}}=0.55$.
Moreover, we can find the largest value $\alpha$ such that the
pair $\{\alpha,e_{\rm{min}}\}$ satisfies both Equation
(\ref{eq:cond_ecc}) and the stability condition
in Equation (\ref{eq:stability}).
We numerically find that the solution is $\alpha=1/7.8$ and 
$e_{\rm{min}}=0.51$. 
In other words, for planetary systems with
initial equal inner and outer eccentricities
and in dynamically stable configurations, the required 
eccentricity and semi-major axis ratio
to reach $e_{{\rm in},f}\to1$ are 
$e_{{\rm in},i}=e_{{\rm out},i}\geq0.51$
and $\alpha<1/7.8$, respectively.

By replacing $e_{{\rm in},i}=e_{{\rm out},i}=0.51$ in the 
angular momentum conservation condition (Eq. [\ref{eq:am}])
we get that the required values of $\mu$ and $\alpha$ 
to reach $e_{{\rm in},f}\to 1$ with the minimum inner
and outer eccentricities are
\ba
\mu \alpha^{1/2}=\frac{1}{(1-0.51^2)^{1/2}}-1=0.162,
\label{eq:mu_alpha2}
\ea
where $\alpha<1/7.8$ and $\mu>0.45$.
In a recent work \citet{petro15b} shows that
the stability condition in Equation (\ref{eq:stability}) is somewhat
conservative and many systems with $\alpha<1/7.8$
are likely to be long-term stable. His results indicate that two Jupiter-mass
planets with eccentricities of 0.5 are long-term stable for 
$\alpha\lesssim1/5$. 
By using these findings by \citet{petro15b}
the conditions on the semi-major axis ratio and 
mass ratio change to  $\alpha\lesssim1/5$ and $\mu\gtrsim0.36$,
while the minimum eccentricities change only slightly from
0.51 to 0.49.

Consistent with our example in Figure \ref{fig:example_sec},
which has a planetary mass ratio of $\mu=2M_J/(3.3M_J)=0.606$
and initial eccentricities of $e_{{\rm in},i}=e_{{\rm in},f}=0.51$, 
we observe from Figure \ref{fig:e_max_mass_e}
that starting from $e_{{\rm in},i}=0.5$ (green line) we
can reach very high eccentricities for $\mu\simeq0.6$ 
and $e_{{\rm out},i}\simeq0.5$.
Similarly, the condition in Equation (\ref{eq:mu_alpha2})
for $\alpha=1/8$ results in $\mu=0.458$, roughly consistent
with the example in Figure \ref{fig:example_sec}.

In summary, the required eccentricity of the outer planet
to excite the eccentricity of the inner planet
up to unity decreases with the initial inner eccentricity
and it reaches a minimum for planetary mass ratios $\mu\sim0.3-1$
when $\alpha=1/10$.
When both planets start with the same eccentricity, the
minimum required initial eccentricities are $\simeq0.5$,
while the dynamical stability of the system requires that
the semi-major axis ratio is $\alpha\lesssim1/5$ and the mass ratio
is $\mu=m_{\rm in}/m_{\rm out}\gtrsim0.36$.

\subsection{Extra forces and maximum
eccentricity growth}
\label{sec:max_e_extra}

We study the effect that extra forces 
have on the three-body system considered
here and how they limit the eccentricity growth.
We do this by including extra terms 
in the orbit-averaged dimensionless potential\footnote{A 
similar approach has been recently and independently 
implemented by \citet{liu14} in the context of 
the Kozai-Lidov mechanism.}  
$\tilde{\phi}_{\rm oct}$ in Equation (\ref{eq:potential}).
For consistency with the positive sign 
in our definition of $\tilde{\phi}_{\rm oct}$,
we also define the interaction potentials as positive 
below. 

The first order general relativistic (GR) correction in the 
planetary approximation ($m_{\rm in},m_{\rm out}\ll m_1$) 
can be written in a dimensionless form as:
\ba
\tilde{\phi}_{\rm{GR}}&=&\frac{\phi_{\rm{GR}}}{\phi_0}=
\frac{4Gm_1^2}{c^2a_{\rm in}m_{\rm in}}\mu \alpha^{-3}
\left(1-e_{\rm{in}}^2\right)^{-1/2},
\label{eq:gr_1}
\ea
where by setting $m_1=1M_\odot$ and $m_{\rm in}=1M_J$, we get
\ba
\tilde{\phi}_{\rm{GR}}&=&
0.0396\left(\frac{1~ {\rm AU}}{a_{\rm in}}\right) \mu 
\left(\frac{0.1}{\alpha}\right)^{3}
\left(1-e_{\rm{in}}^2\right)^{-1/2}.
\label{eq:gr_2}
\ea

Similarly, the dimensionless potential due to the tidal 
quadrupole on the planet can be written as 
(e.g., \citealt{FT07})
\ba
\tilde{\phi}_{\rm{tq}}=\frac{\phi_{\rm{tidal}}}{\phi_0}&=&
\frac{4k_p}{3} \left(\frac{m_1}{m_{\rm in}}\right)^2
\left(\frac{R_{\rm in}}{a_{\rm in}}\right)^{5}
\mu \alpha^{-3}\nonumber\\
&&\times\frac{1+3e_{\rm{in}}^2+3e_{\rm{in}}^4/8}{(1-e_{\rm{in}}^2)^{9/2}},
\label{eq:tide_1}
\ea
where for $m_1=1M_\odot$, $m_{\rm in}=1M_J$, a tidal
Love number of the planet of $k_p=0.26$, and 
radius of the inner planet $R_{\rm in}=R_J$, we get
\ba
\tilde{\phi}_{\rm{tq}}=\frac{\phi_{\rm{tidal}}}{\phi_0}&=&
7.93\times10^{-9}\left(\frac{1\mbox{ AU}}{a_{\rm in}}\right)^{5}\mu 
\left(\frac{0.1}{\alpha}\right)^{3}\nonumber\\
&&\times\frac{1+3e_{\rm{in}}^2+3e_{\rm{in}}^4/8}{(1-e_{\rm{in}}^2)^{9/2}}.
\label{eq:tide_2}
\ea

We note that with these parameters 
both GR and tidal quadrupole contributions 
can become comparable to $\tilde\phi_{\rm oct}$ in Equation 
(\ref{eq:potential}), which 
is of order unity,  only at very high eccentricities or small
semi-major axis $a_{\rm in}$. 
For the parameters in Equations (\ref{eq:gr_2}) and (\ref{eq:tide_2}) we
get that $\tilde{\phi}_{\rm{GR}}=\tilde{\phi}_{\rm{tq}}$ at 
eccentricities of $e_{\rm in}\simeq0.983$.
For $e_{\rm in}<0.983$ GR dominates over the tidal bulge, while
the opposite happens for $e_{\rm in}>0.983$.

We write the dimensionless potential that includes the extra 
forces as:
\ba
\tilde{\phi}_{\rm{extra}}\equiv\tilde{\phi}_{\rm{oct}}+
\tilde{\phi}_{\rm{GR}}+\tilde{\phi}_{\rm{tq}}.
\ea

In Figure \ref{fig:emax_phi} we show the maximum eccentricity 
of the inner planet as a function of $e_{{\rm out},i}$
and the mass ratio $\mu=m_{\rm in}/m_{\rm out}$ from  solving the equation:
\ba 
\tilde{\phi}_{\rm{extra}}(e_{\rm in,i},e_{\rm out,i},\varpi_i=\pi)
=\tilde{\phi}_{\rm{extra}}(e_{\rm in,f},e_{\rm out,f},\varpi_f),
\label{eq:cond_full_3}
\ea
where we fix the initial eccentricity of the inner planet to 
$e_{{\rm in},i}=0.6$ and the masses of the star and the inner
planet to $m_1=1M_\odot$ and $m_{\rm in}=1M_J$, respectively.
By using the total angular momentum conservation (Eq. [\ref{eq:am}])
we can solve for $e_{{\rm in},f}$ (or $e_{{\rm out},f}$)
and $\varpi_f$. The maximum eccentricity of the
inner planet is obtained by numerically maximizing over 
$\varpi_{f}$.

In panel a, we show our results for  $a_{\rm in}=1$ AU
and $a_{\rm out}=10$ AU.
We observe that the maximum eccentricity is limited 
(i.e., $e_{{\rm in},f}<1$) by the inclusion of the extra forces. 
For comparison we show the minimum 
$e_{{\rm out},i}$ required to reach  $e_{{\rm in},f}=1$
(similar to Figure \ref{fig:e_max_mass_e})
when no extra forces are included
(dot-dashed gray line).
We observe that the extra forces limit 
the maximum eccentricity more efficiently for larger values
of $\mu=m_{\rm in}/m_{\rm out}$. For instance, 
for $\mu=0.01$ (0.4) we get that $e_{{\rm in},f}=1$ with no extra
forces and a minimum $e_{{\rm out},i}\simeq0.62$ 
($\simeq0.47$ ), while for the same value of $e_{{\rm out},i}$ 
the extra forces yield a maximum eccentricity of 
$\simeq0.99$ ($<0.95$).
This is because for more massive outer perturbers both 
$\tilde{\phi}_{\rm{GR}}$ and $\tilde{\phi}_{\rm{tq}}$
are smaller: the extra forces do not depend on the mass
of the perturber, while the point-like gravitational interactions
increase linearly in magnitude with $m_{\rm out}$.

From panel a, we note that the maximum eccentricity the inner orbit
can reach is always less than $\simeq 0.985$ for dynamically 
stable  configurations (left of the stability boundary, 
dashed black line).
This result implies that the pericenter distance is 
$r_p\equiv a_{\rm in}(1-e_{\rm in})>0.015$ AU and, therefore, no
tidal disruptions are expected for these parameters.
Moreover, if the planets undergo migration at roughly
constant  angular momentum then their
final semi-major axis is roughly twice the minimum
pericenter distance, which implies that the semi-major
axes of the hot Jupiters are constrained to $a>0.03$ AU.

In panel b, we show our results for  $a_{\rm in}=5$ AU
and $a_{\rm out}=50$ AU.
We observe that the 
maximum eccentricity is higher than that with 
$a_{\rm in}=1$ AU (panel a), which is expected
because both $\tilde{\phi}_{\rm{GR}}$ and $\tilde{\phi}_{\rm{tq}}$
decrease with $a_{\rm in}$, while  $\tilde{\phi}_{\rm{oct}}$ remains
constant (at fixed $\alpha$). 
In particular, we observe that a large fraction of the area 
displayed in the plot reaches a maximum eccentricity of 
$\simeq0.995$ (in dark red).

By only considering that GR as an extra force, 
we can roughly estimate the dependence of the maximum 
eccentricity on $a_{\rm in}$ from Equation (\ref{eq:cond_full_3})
by using that in the initial state 
$\tilde{\phi}_{\rm{GR}}\ll\tilde{\phi}_{\rm{oct}}$  
(initial eccentricity is not too high or $a_{\rm in}$ is not too small) 
and that in the final state 
with $1-e_{\rm in,f}\ll 1$, $\tilde{\phi}_{\rm{oct}}$ is approximately 
independent on the eccentricity.
Thus, from Equation (\ref{eq:cond_full_3}) and only varying 
$a_{\rm in}$ ($\mu$) and $e_{\rm in}$, we get that the maximum eccentricity
in the final state depends on the semi-major axis (mass ratio $\mu$) as
$1-e_{\rm in,max}\propto a_{\rm in}^{-2}$ 
($1-e_{\rm in,max}\propto \mu^{2}$)
A similar reasoning yields  a scaling 
$1-e_{\rm in,max}\propto a_{\rm in}^{-10/9}$ (and 
$1-e_{\rm in,max}\propto \mu^{2/9}$) if the dominant 
extra force is the tidal quadrupole.

Despite the larger eccentricities observed with 
$a_{\rm in}=5$ AU compared to $a_{\rm in}=1$ AU, the
minimum pericenter distances in dynamically stable
configurations are similar.
For these configurations (left of the black dashed line)
we have that  $e_{\rm in}<0.9974$ for $a_{\rm in}=5$ AU,
which implies that the pericenter distance
is  $r_p=a_{\rm in}(1-e_{\rm in})>0.013$ AU, compared to
  $r_p=a_{\rm in}(1-e_{\rm in})>0.015$ AU for  $a_{\rm in}=1$ 
  AU.
This is consistent with the dependence of the maximum 
eccentricity on $a_{\rm in}$ given above, which would 
translate in a minimum pericenter distance 
$r_{p, \rm min}=a_{\rm in}(1-e_{\rm in,max})$ that goes like
$r_{p, \rm min}=a_{\rm in}^{-1}$ and 
$r_{p, \rm min}=a_{\rm in}^{-1/9}$ if GR and the tidal quadrupole
dominates, respectively. 
Then, since the tidal quadrupole dominates in this regime 
of extreme eccentricities
($\tilde{\phi}_{\rm{GR}}<\tilde{\phi}_{\rm{tq}}$ for
$e_{\rm in}<0.983$ and $a_{\rm in}=1$ AU
from  Eqs. [\ref{eq:gr_2}] and [\ref{eq:tide_2}]) we 
expect very little dependence of the minimum pericenter
on $a_{\rm in}$.
  
Finally, we have only studied a limited part of the
phase-space and there are additional parameters that could 
be varied. 
Probably the most relevant is the semi-major axis ratio
$\alpha$. We experimented by repeating 
panels a and b with $\alpha$ reduced from
$1/10$ to $1/20$ and  found that the maximum eccentricities 
are reduced, which is expected since the gravitational secular 
interactions from $\tilde{\phi}_{\rm{oct}}$ become weaker.

In summary, adding GR and tidal quadrupole terms
to the three-body Newtonian point-like gravitational interactions limits 
the maximum eccentricity (or minimum pericenter distance). 
This effect has two important 
consequences: the planets generally avoid being tidally disrupted  
and the hot Jupiters formed by this mechanism have a 
minimum semi-major axis of $\sim0.03$ AU.

\subsection{Departure from coplanarity}

Our analysis above assumes that the inner 
and outer orbits are coplanar ($i_{\rm tot}=0$).
This limit should be a good approximation for small
departures from coplanarity since the dynamics is described
 by the potential $\tilde{\phi}_{\rm{oct}}$ (Eq. [\ref{eq:potential}]), which
 is accurate to first order
in the mutual inclination $i_{\rm tot}$.

We have empirically found that CHEM operates roughly as described
by our analytical analysis when $i_{\rm tot}\lesssim20^\circ$.
In particular, we have varied the mutual inclination 
using the secular evolution equations from \citet{petro15} and
checked in a few cases that the eccentricity of the inner orbit 
reaches $e_{\rm in}\simeq1$ starting from $e_{\rm in}\simeq0$
and $e_{\rm out}$ from Figure \ref{fig:e_max_mass} 
when $i_{\rm tot}\lesssim20^\circ$.
For $i_{\rm tot}\sim20-50^\circ$ there are
eccentricity oscillations that occur in the quadrupole timescale
that tend to limit the eccentricity growth and the description by
our analytical theory becomes poor. 
For large enough mutual inclinations ($i_{\rm tot}\gtrsim60^\circ$)
the inner eccentricity tend to reach $\simeq1$ by the Kozai-Lidov
mechanism (see \citealt{teyss13} for a systematic 
study of this regime).

From our limited exploration of parameters and initial
orbital configurations we note that CHEM is not necessarily 
quenched by considering somewhat large initial mutual inclinations 
($i_{\rm tot}\gtrsim20^\circ$), but the description of the eccentricity
forcing changes in nature and is dominated by quadrupole 
timescale (see  \citealt{li14b} for an exploration
of this regime in the test particle approximation). 
A systematic parameter survey in the non-coplanar
regime is beyond the scope of this paper.

\begin{figure*}
   \centering
  \includegraphics[width=18cm]{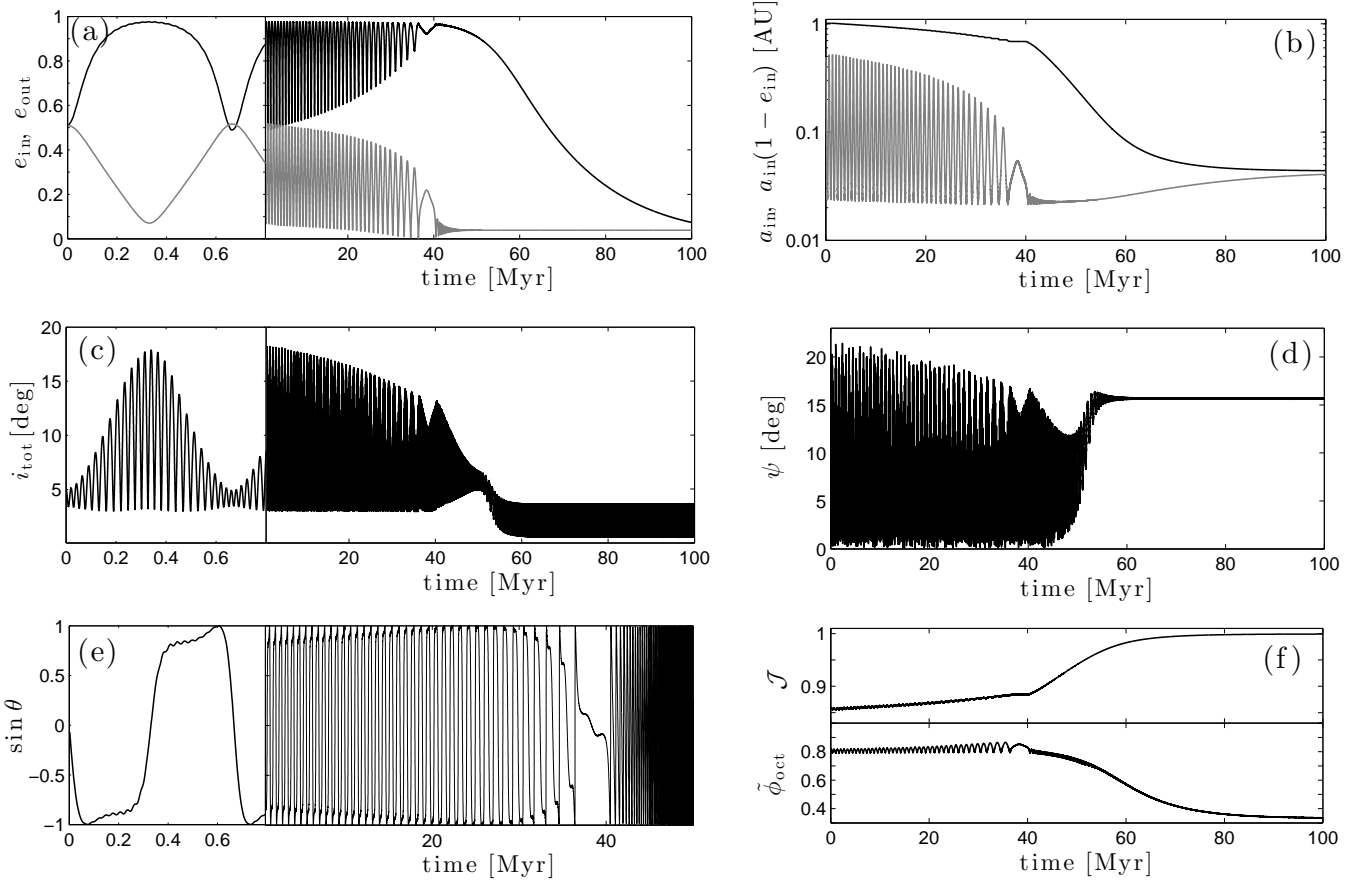} 
  \caption{Evolution of two planets 
  initially in orbits (inner and outer)  
  with  $a_{\rm in}=1$ AU, $e_{\rm in}=0.51$, 
  $a_{\rm out}=8$ AU, $e_{\rm out}=0.51$,
  and mutual inclination  $i_{\rm tot}=5^\circ$. 
  Initially the arguments of pericenter are 
  $\omega_{\rm in}=\omega_{\rm out}=0$,
 the ascending nodes are
 $\Omega_{\rm in}=0$ and  $\Omega_{\rm out}=180^\circ$,
 and the stellar obliquity is $\psi=0$ (angle between 
the stellar spin and angular momentum vector of
the inner planet).
  The planets have masses of
  $m_{\rm in}=2M_J$  and   $m_{\rm out}=3.3M_J$.
  We chose planetary and stellar viscous times of 
  $t_{V,p}=0.03$ yr 
  and $t_{V,s}=50$ yr, respectively (see text in \S\ref{sec:evolution}). 
Panel (a): eccentricities of the inner ($e_{\rm in}$, black line)
and outer ($e_{\rm out}$, gray line) planets.
Panel (b): semi-major axis (black line) and pericenter 
distance $a_{\rm in}(1-e_{\rm in})$ (gray line) of
the inner planet.
Panel (c): mutual inclination between the two
planetary orbits.
Panel (d): stellar obliquity (angle between the host star's spin
axis and the orbital angular momentum vector of the inner orbit).
Panel (e): 
$\sin \theta=\left( \bf{\hat{e}}_{\rm out} 
\times \bf{\hat{e}}_{\rm in}\right)\cdot\bf{\hat{h}}_{\rm in}$ 
(Eq. [\ref{eq:sin_theta}]). Note that the maximum time shown
in this panel is $50$ Myr, as opposed to 100 Myr as in the other panels.
Panel (f): dimensionless total orbital 
angular momentum $\mathcal{J}$ in Equation (\ref{eq:am})
(upper panel) and dimensionless interaction potential 
$\tilde{\phi}_{\rm oct}$ in Equation (\ref{eq:potential}) (lower panel).
}
\label{fig:example_sec}
\end{figure*}  

\section{Evolution during Migration}
\label{sec:evolution}

In Figure \ref{fig:example_sec}, we show an example of 
the secular evolution of two planets in initially eccentric 
($e_{\rm in}=e_{\rm out}=0.51$) and low mutual inclination 
($i_{\rm tot}=5^\circ$) orbits . 

The equations of motion are fully described in  \citet{petro15} (see
appendix A therein), where   the efficiency of tidal dissipation
is parametrized by the viscous timescales of the star
and the planet $t_{V,s}$ and $t_{V,p}$.
For reference, a highly eccentric  ($1-e\ll 1$) Jupiter-like 
planet orbiting a Solar-mass star with $a>1$ AU can 
be circularized
to become a hot Jupiter with final $a=0.05$ AU
($a=0.06$ AU) within 1 Gyr for $t_{V,p}\lesssim0.35$ yr 
 ($t_{V,p}\lesssim0.15$ yr) \citep{SKD12}. Given our choice of 
 $t_{V,s}$ and $t_{V,p}$, tides in the planet dominate the
 circularization of the planetary orbit.


From panel a, we observe that the inner and outer planets
efficiently exchange angular momentum:
the inner orbit oscillates in eccentricity in the range 
$\simeq0.48-0.97$, while the outer orbit does so in 
the range $\simeq0.05-0.52$.
These large-amplitude oscillations allow the inner planet to reach a 
minimum pericenter distance of 
$a_{\rm in}(1-e_{\rm in})\simeq0.024$ AU where
tidal dissipation can efficiently extract orbital energy (panel b).
Thus, the orbit shrinks steadily during the phases  
in which the pericenter distances are small.
From panel b, we observe that the semi-major axis decays almost 
linearly during the first $\sim40$ Myr, after which the eccentricity
oscillations are damped and the migration speeds up.
The final semi-major axis of the HJ formed in this example is 
$\simeq0.044$ AU, which roughly corresponds to the mean
and median of $\simeq0.05$ AU observed population of 
hot Jupiters detected in RV and transit surveys.

From panel c, we observe that the mutual inclination between
the planetary orbits oscillates in the range $i_{\rm tot}\simeq3-18^\circ$. 
The time at which  $i_{\rm tot}$ reaches its maximum value of 
$\simeq 18^\circ$ coincides  
with the time at which $e_{\rm in}$ also reaches a maximum. 
However, the inclination shows many oscillations within one
oscillation of the eccentricities because the former varies in
the quadrupole timescale, while the latter does so in
the octupole timescale \citep{li14}.
We note that in the coplanar limit ($i_{\rm tot}=0$), 
the quadrupole potential is axisymmetric 
(see first term in Equation [\ref{eq:potential}]), implying that
it does not drive any angular momentum exchange 
between the orbits. 
However, if the orbits have non-zero (but still small) mutual 
inclinations, the quadrupole potential can still drive small
amplitude eccentricity and inclination oscillations. 
Given the scale of panel c relative to that in panel a, 
the quadrupole-driven 
oscillations  can only be observed in $i_{\rm tot}$ and
their amplitude is modulated
by the octupole potential.

Once the eccentricity oscillations are damped at $\sim40$ Myr the 
inclination oscillations are no longer modulated by the octupole 
and vary only in the quadrupole timescale with decreasing amplitude.
These oscillations change in character at $\gtrsim50$ Myr, after which  
the mutual inclination damps to small values ($i_{\rm tot}\lesssim3^\circ$)
and oscillates due to the
planetary orbital precession produced by the host star's bulge.
This flattening of the inner orbit has been previously observed by  
\citet{correia13} in a similar context of hierarchical two-planet
systems.

From panel d we observe that the stellar obliquity 
(i.e., the angle between the host star's spin axis
and the orbital angular momentum vector of the inner orbit) 
starts oscillating in the range $\psi\simeq 0-20^\circ$ due to the perturbations 
of the outer planet. 
Once the planetary orbit starts flattening at $\gtrsim50$ Myr, the 
conservation  of angular momentum forces the obliquity to 
increase and it does so from $\sim10^\circ$ to $\sim20^\circ$.
After $\gtrsim60$ Myr the semi-major axis is $\lesssim0.2$ AU
and the planetary orbital precession is dominated by host star's
bulge rather than the outer planet. Thus, the stellar obliquity 
stabilizes\footnote{
We note that depending on the efficiency of 
the tidal dissipation in the star (i.e., the stellar viscous time
$t_{V,s}$), the value of $\psi$ and $i_{\rm tot}$ could still 
change after the planetary orbit has circularized.}
 at $\psi\simeq16^\circ$.

In summary, our example ends with the formation of a hot 
Jupiter at $a\simeq0.044$ AU with a stellar obliquity of 
$\psi\simeq16^\circ$
and planetary perturber at 8 AU, which is in a 
circular and nearly coplanar orbit relative to the hot Jupiter.

\subsection{Secular eccentricity forcing}

The equations of motion of the eccentricity and
angular momentum vectors can be obtained by taking 
gradients of the dimensionless potential 
(e.g., \citealt{tremaine09,petro15}).
Evidently, from Equation (\ref{eq:potential}) 
$\nabla_{h_{\rm{in}}} \tilde{\phi}_{\rm oct}=0$
and the equation of motion of the eccentricity 
vector of the inner planet becomes
\ba
\frac{d {\bf e}_{\rm{in}}}{dt} =\frac{1}{\tau_{\rm in}} 
(1-e_{\rm{in}}^2)^{1/2} {\bf \hat{h}}_{\rm{in}} \times 
\nabla_{e_{\rm{in}}} \tilde{\phi}_{\rm oct},
\label{eq:de_dt}
\ea
where
\ba
 \tau_{\rm in} =\frac{2}{3\pi}\left(\frac{m_1}{m_{\rm out}}\right)
\left(\frac{a_{\rm out}}{a_{\rm in}}\right)^3P_{\rm in}
\label{eq:tau_in}
\ea
and $P_{\rm in}$ is the orbital period of the inner planet.
From this equation, we can 
calculate the eccentricity forcing term (i.e., the term 
proportional to ${\bf \hat{e}}_{\rm{in}}$
in Eq. [\ref{eq:de_dt}])
as (see also \citealt{LP03,li14})
\ba
\frac{de_{\rm in}}{dt}\simeq-\frac{5\alpha}{4\tau_{\rm in}}
\frac{e_{\rm out}(1-e_{\rm in}^2)^{1/2}
(1+3e_{\rm in}^2/4) }{(1-e_{\rm out}^2)^{5/2}}\sin\theta,
\label{eq:e_dot}
\ea
where we define
\ba
\sin \theta=\left( \bf{\hat{e}}_{\rm out} \times \bf{\hat{e}}_{\rm in}\right)
\cdot\bf{\hat{h}}_{\rm in}.
\label{eq:sin_theta}
\ea
Note that the angle $\theta$ coincides with 
$\varpi=\varpi_{\rm in}-\varpi_{\rm out}$
when the orbits are coplanar.

Similarly, one can differentiate 
${\bf e}_{\rm in}\cdot{\bf e}_{\rm out}\equiv 
e_{\rm in}e_{\rm out}\cos\varpi$ to 
find\footnote{Note that 
$d {\bf e}_{\rm{out}}/dt =\tau_{\rm out}^{-1}
(1-e_{\rm{out}}^2)^{1/2} {\bf \hat{h}}_{\rm{in}} \times 
\nabla_{e_{\rm{out}}} \tilde{\phi}_{\rm oct}$ and 
$\tau_{\rm in}/\tau_{\rm out}=\mu\alpha^{1/2}$.}
\ba
\frac{d\varpi}{dt}&\simeq&\frac{1}{\tau_{\rm in}}
\bigg\{
\frac{(1-e_{\rm in}^2)^{1/2}}{(1-e_{\rm out}^2)^{3/2}} -
\mu\alpha^{1/2}\frac{(1+3e_{\rm in}^2/2) }{(1-e_{\rm out}^2)^{2}}\nonumber\\
&&-\frac{5\alpha}{4}\bigg[\frac{e_{\rm out}(1-e_{\rm in}^2)^{1/2}
(1+9e_{\rm in}^2/4)}{e_{\rm in}(1-e_{\rm out}^2)^{5/2}}\nonumber\\
&&-\mu\alpha^{1/2}\frac{e_{\rm in}(1+4e_{\rm out}^2)
(1+3e_{\rm in}^2/4)}{e_{\rm out}(1-e_{\rm out}^2)^{3}}
\bigg]\cos\varpi\bigg\}.
\label{eq:omega_dot}
\ea

 We note from Equations (\ref{eq:e_dot}) and (\ref{eq:omega_dot})
that the planet masses change the 
timescale of the secular gravitational interactions through $\tau_{\rm in}$
and the evolution of $\varpi$ through $\mu$.

The timescale for the eccentricity growth is roughly
 $\tau_{\rm in}/\alpha$, which for our example in 
 Figure \ref{fig:example_sec} it corresponds to $\simeq0.27$ Myr,
 consistent with the timescale of $\sim0.3$ Myr 
 that it takes for the eccentricity to grow from $\simeq0.5$ to 
$\simeq 1$ (panel a).

In panel e of Figure \ref{fig:example_sec}, we show the evolution of 
$\sin \theta$ for the example. 
As expected from Equation (\ref{eq:e_dot}), we observe 
that the eccentricity of the inner planet
(panel a)  increases (decreases) when 
$\sin \theta<0$ ($\sin \theta>0$).

In this simulation $\sin\theta$ starts at $0$  and 
rapidly decreases to $\simeq-1$ ($\theta\simeq270^\circ$), 
where it remains oscillating close to this value.
At some point,  $\sin\theta$ jumps from  $\simeq-1$ to 
$\simeq1$ ($\theta\simeq90^\circ$) and stays around
this angle, while the eccentricity of the inner planet
starts decreasing. 
This behavior is sketched in the energy levels of
Figure \ref{fig:contour} (black line in panel c), where we
observe that the eccentricity growth (or decrease) 
happens mostly for $\varpi\sim90^\circ$ 
(or $\sim270^\circ$).

This behavior can be understood from Equation 
(\ref{eq:e_dot}) where we observe that the slow 
variation of $\sin\theta$ around $\pm1$ allows
for a persistent eccentricity growth or decay.
Moreover, from Equation (\ref{eq:omega_dot}) the slow variation of 
$\sin\theta$  around $\pm1$ (i.e., $\cos\varpi\simeq0$) 
happens when $e_{\rm in}$ and $e_{\rm out}$ are such that
$\mu\alpha^{1/2}\simeq(1-e_{\rm in}^2)^{1/2}(1-e_{\rm out}^2)^{1/2}
/(1+3e_{\rm in}^2/2)$, so $d\varpi/dt\simeq0$
\citep{LP03}.
This last condition implies that in the test particle approximation
($\mu\ll1$), this resonant-like behavior can not happen unless
$1-e_{\rm in}\ll1$.

In summary, in this example we show that the eccentricity forcing
can be enhanced by having a slow variation
of $\varpi$ around $90^\circ$ or $270^\circ$, which 
can achieved for either not too small values of $\mu\alpha^{1/2}$ 
or high enough eccentricities.

\begin{figure}
   \centering
  \includegraphics[width=8.3cm]{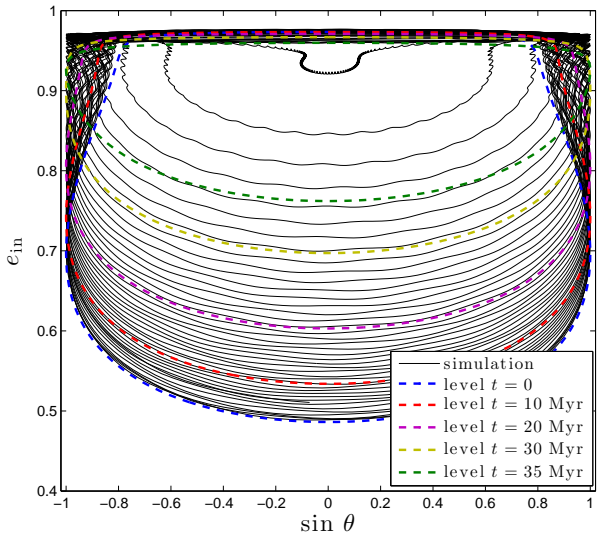} 
  \caption{Evolution of $\sin\theta$ and the
  inner eccentricity from the example in Figure \ref{fig:example_sec}
  for the first 42.4 Myr of the simulation, at which time the
  oscillations are fully quenched. We also plot
  the contour levels of $\tilde{\phi}_{\rm extra}$ in Equation 
  (\ref{eq:cond_full_3}) using the values of $\alpha$, $\mathcal{J}$,
  and $\tilde{\phi}_{\rm extra}$ that correspond to different times
  of the example, as labeled.
  }
\label{fig:e_sin}
\end{figure}

\begin{figure*}
   \centering
  \includegraphics[width=18cm]{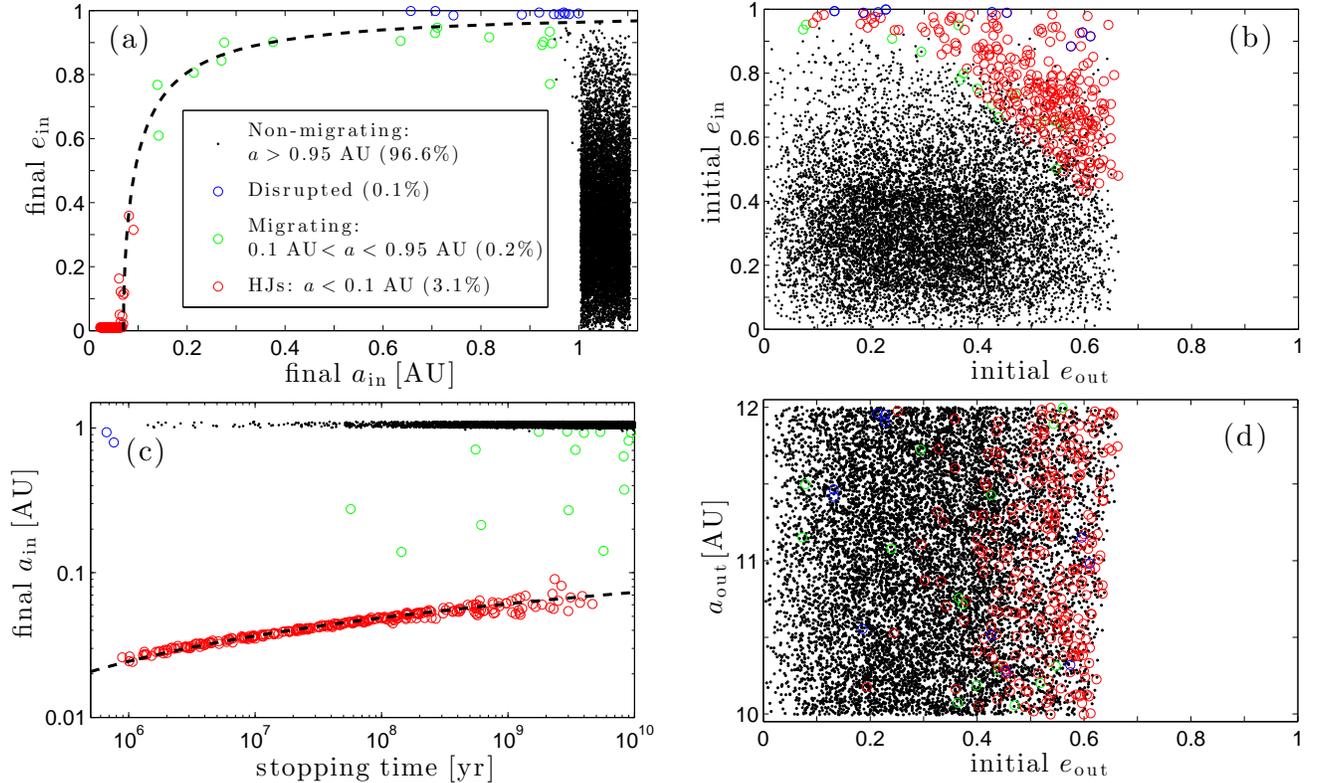}
  \caption{Outcomes for our population synthesis study, 
  as labeled in panel (a).
  We chose planetary and stellar viscous times of 
  $t_{V,p}=0.1$ yr 
  and $t_{V,s}=50$ yr, respectively (see text in \S\ref{sec:evolution}). 
The inner planet has a mass of $1M_J$ and an initial semi-major axis
drawn from a uniform distribution in [1,1.1] AU, while the outer planet 
has a mass uniformly distributed in $[1.3,1.7] M_J$
and an initial semi-major axis drawn from a uniform distribution
in $[10,12]$ AU.
The eccentricities of both the inner and outer planets are drawn from the 
distribution in Equation (\ref{eq:sigma_e}) with $\sigma_e=0.3$.
  Panel (a): final semi-major axis versus final eccentricity of the planetary orbit.
  The constant angular momentum track $a_{\rm in}(1-e_{\rm in}^2)=0.07$ AU
  is indicated by a dashed line.
  Panel (b): initial eccentricity of the outer planet $e_{\rm out}$
 versus the initial eccentricity of the inner planet $e_{\rm in}$.
 The boundary at high eccentricity reflects the stability condition
for hierarchical triple systems 
given by Equation (\ref{eq:stability}).
   Panel (c):  time at which the simulation is stopped 
   versus the final semi-major axis of the inner planet
   $a_{\rm in}$. The dashed line shows the empirical fit 
   $\log\left( t_{\rm mig}/1 {\rm Myr}\right)= 0.82\times a_{\rm in}/0.01~{\rm AU}-2$, where
   $t_{\rm mig}$ is the timescale for the migration of the HJ with 
   a final semi-major   axis $a_{\rm in}$. 
Panel (d):  initial eccentricity $e_{\rm out}$ versus
the semi-major axis of the outer planet.
\\
\\
}
\label{fig:montecarlo}
\end{figure*}  

\subsection{Quenching of the eccentricity 
oscillations}

We observe from Figure \ref{fig:example_sec} that
as the semi-major axis shrinks (panel b), the
eccentricity oscillations of the inner planet 
start to damp: the minimum value of $e_{\rm in}$
in each oscillation increases as a function
of time.
The oscillations are completely damped at $\simeq42$
Myr.

We observe from panel e that the oscillatory 
behavior of $\sin\theta$ discussed in the previous 
section continues up to $\sim30$ Myr and then 
the planet gradually starts spending less time 
at $\sin\theta\sim1$, where the eccentricity
forcing is maximum. Then, at $\sim40$
Myr $\sin\theta$ stops librating and circulates
in a timescale that is shorter than the
octupole timescale.

In panel f we show the evolution of the
dimensionless angular momentum $\mathcal{J}$
(Eq. [\ref{eq:am}]) and potential $\tilde{\phi}_{\rm oct}$
(Eq. [\ref{eq:potential}]).

First, we observe that $\mathcal{J}$ increases 
nearly monotonically from $\simeq0.86$ to
$1$ (i.e., to two nearly circular orbits). 
Second, $\tilde{\phi}_{\rm oct}$ stays roughly constant
with small oscillations around $\simeq0.83$
during the first $\sim40$ Myr and then decreases monotonically
to $\simeq1/3$ (i.e., $e_{\rm in}=e_{\rm in}=0$ in 
Eq. [\ref{eq:potential}]).

In Figure \ref{fig:e_sin} we show the evolution of 
$e_{\rm in}$ and $\sin\theta$ from our example.
We show the results up to a maximum time of 
$42.4$ Myr, at which time the eccentricity oscillations
are almost fully quenched and $\theta$ starts 
circulating. 
We also plot the phase-space trajectories from the energy 
contours of $\tilde{\phi}_{\rm extra}$ in Equation 
  (\ref{eq:cond_full_3}) and fixing $a_{\rm in}$, $\mathcal{J}$,
and $\tilde{\phi}_{\rm extra}$ to match the simulation
at different times. We observe that the phase-space 
trajectories roughly match the numerical example
and describe well the quenching of the eccentricity
oscillations.
This result shows that the quenching of the eccentricity
oscillations is mainly due to the 
monotonic increase of $\mathcal{J}$ in time. 

Thus, this analysis suggests that in order to have eccentricity 
oscillations down to smaller $a_{\rm in}$ (or smaller $\alpha$) during
migration, one might either need to start from smaller $\mathcal{J}$
or set $\mu$ to be smaller so $\mathcal{J}$ increases more 
slowly with the decreasing $\alpha$.

\section{Population synthesis study}

We ran a series of numerical experiments to study
 the evolution of triple 
systems consisting a sun-like host star ($m_1=1 M_\odot$
and $R_1=R _\odot$) and two orbiting
planets with masses $m_{\rm in}$ and $m_{\rm out}$.
The inner planet has $m_{\rm in}=1M_J$ and Jupiter radius, while
the outer has a mass that is randomly distributed in 
$[1.3,1.7]M_J$. 
 This choice of masses is motivated by our results in
\S\ref{sec:mig_space}, where we find that 
CHEM works best for outer planets slightly more massive than 
the inner planet.
The equations of motion are fully described in \citet{petro15}.

The initial eccentricity and mutual inclination of the planets follow a 
Rayleigh distribution:
\ba
dp=\frac{x\,dx}{\sigma_x^2} \exp\left(-\frac{1}{2} x^2/\sigma_x^2\right),
\label{eq:sigma_e}
\ea
where $x=i,e$.
We choose $\sigma_e=0.3$, which is intended to represent 
the tail\footnote{CHEM mostly works for $e>0.3$
so we do not attempt to model the eccentricity
distribution for lower eccentricities.} ($e\gtrsim0.3$)
of the observed eccentricity distribution of 
giant planets ($m\sin i>0.1 M_J$) with periods longer than 1 year.
For the mutual inclinations we choose $\sigma_i=0.1$, or a mean
of $\approx 7^\circ$, which is slightly higher than the upper
limit to the mean mutual inclination of $\approx 5^\circ$
constrained from Kepler  \citep{TD12,fab14}.

The semi-major axis of the inner planet is 
drawn from a uniform distribution in $[1,1.1]$ AU,
while that of the outer planet is drawn from a
uniform distribution in $[10,12]$ AU.
We discard systems that do not satisfy the stability condition
\citep{MA01}:
\ba
\frac{a_{\rm{out}}}{a_{\rm{in}}}&>& 
2.8(1+\tilde{\mu})^{2/5}\frac{(1+e_{\rm out})^{2/5}}{(1-e_{\rm out})^{6/5}}  
\left(1-0.3\frac{i_{\rm tot}}{180^\circ}\right)
\label{eq:stability}
\ea
where $\tilde{\mu}=m_{\rm out}/(m_1+m_{\rm in})$.

The longitudes of the arguments of pericenter and 
longitude of the ascending node are chosen
randomly  for the inner and outer orbits. 
The host star and the planet start spinning with 
periods of 10 days and 10 hours, respectively,
both along the ${\bf \hat{h}}_{\rm{in},0}$ axis, 
implying that the initial obliquities are zero.

Finally, we stop each run when a maximum time 
chosen uniformly in $[0,10]$ Gyr has passed or 
when either a hot Jupiter in a circular orbit 
($e_{\rm in }<0.01$) is formed or a planet is 
tidally disrupted, which we define to occur
when the pericenter distance 
is less than 0.0127 AU \citep{GRL11}.

\subsection{Results}

In Figure \ref{fig:montecarlo}, we show the results 
from our population synthesis study, 
which consists of 9,000 systems.

Most systems ($\simeq96.6\%$, black dots) do not reach eccentricities 
that are high enough to allow for migration. 
In these systems, the mean eccentricity of the inner planet increases only
slightly from an initial value of $\simeq0.34$ to a final value of
$\simeq0.35$. 
Actually, the steady-state final eccentricity distribution looks 
essentially identical to the initial distribution, which means that 
by  construction it can reproduce the observed eccentricity 
distribution of planets at $>1$ AU.

The second most common outcome ($\simeq 3.1\%$) is a system 
with a hot Jupiter ($a_{\rm in}<0.1$ AU, red circles).
From panel b we observe that these systems initially have 
large eccentricities: the mean eccentricity of the inner
and outer planets is 0.71 and 0.52, respectively.
Note that the maximum eccentricity of the outer planet is
$\simeq0.66$, which is an artifact of the stability criterion 
in Equation  (\ref{eq:stability}) (see boundary at  high
$e_{{\rm out},i}$ in panel d of Figure \ref{fig:montecarlo}).
As discussed in \S\ref{sec:analytic},
in order to form a hot Jupiter from an initial circular orbit we
require a perturber with $e_{\rm out}>0.67$, which
explains the lack of hot Jupiters that come from initial
eccentricities $e_{\rm in}<0.4$.
This restriction can be relaxed by using a less restrictive 
stability boundary for hierarchical triple systems
like the ones proposed by \citet{EK95} and
\citet{petro15b}.

The third most common outcome ($\simeq0.2\%$) is a system with a 
migrating planet (0.1 AU $<a_{\rm in}<0.95$ AU, green circles).
From panel a we observe that these systems have high eccentricities
close to the angular momentum track $a_{\rm in}(1-e_{\rm in}^2)=0.07$ AU.

Finally, the least common outcome ($\simeq0.1\%$) is a system
in which the inner planet gets tidally disrupted  
($a_{\rm in}[1-e_{\rm in}]<0.0127$ AU at some point of the simulation,
blue circles). 
Most of these systems start from very high eccentricities
($e_{\rm in}>0.99$) and crossed 
the tidal disruption boundary at the start of the
simulation.

\begin{figure}
   \centering
  \includegraphics[width=8.8cm]{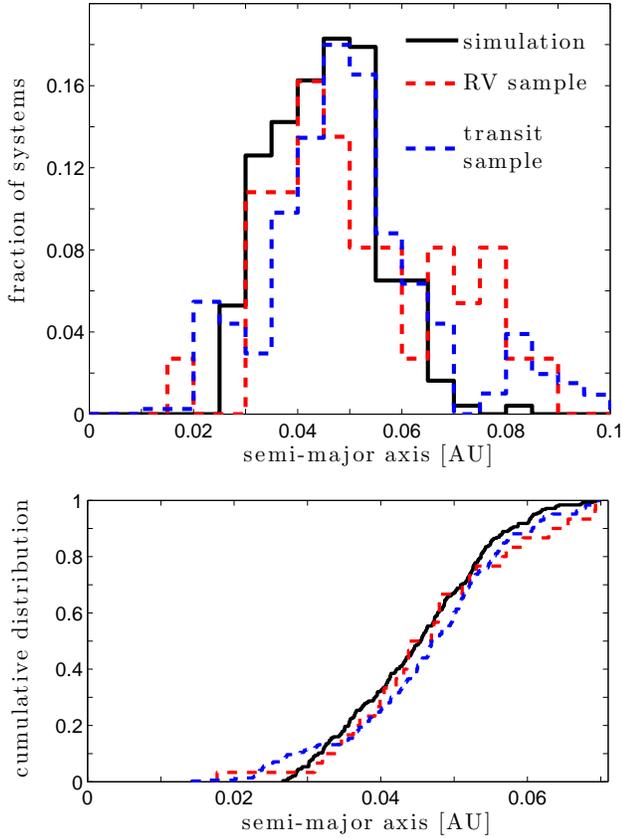} 
  \caption{Semi-major axis distribution of the hot
  Jupiters formed in our population synthesis
  study (solid line) and the observed distribution 
  of planets with $m\sin i>0.1M_J$ detected in RV (red dashed
  line) and transit (blue dashed line) surveys.
  The transit sample is corrected for the geometric selection
  bias.
 The cumulative distribution in the lower panel is
  restricted to hot Jupiters with semi-major
  axis up to 0.07 AU.
  }
\label{fig:hj_sma}
\end{figure}  

\subsection{Semi-major axis distribution
of hot Jupiters}
\label{sec:hj_sma}

In Figure \ref{fig:hj_sma} we show the 
semi-major axis distribution for the hot Jupiters formed in our 
population synthesis study and the observations of hot Jupiters 
with $m\sin i>0.1M_J$ detected in the transit and 
RV surveys\footnote{From The Exoplanet Orbit Database 
\citep{wright11} }.

From the upper panel, we observe that the distribution of 
semi-major axis in the simulation roughly
matches the peak of the observed distribution: the mean
(median) in the simulation are $\simeq0.045$ AU 
($\simeq0.045$ AU), while the observations have 
0.052 AU (0.048 AU) and 0.050 AU (0.049 AU) in the RV and
transit\footnote{The transit sample is corrected by the geometric 
selection bias only.} samples, respectively.

We note that the semi-major axis distribution drops for
$a\lesssim0.03$ AU, which is consistent with our analysis in
\S2.3 where we show that the minimum pericenter distance
that this mechanism can achieve (for similar parameters)
is $\simeq0.015$ AU implying a minimum semi-major
axis the HJs of   $\simeq0.03$ AU\footnote{The
orbital angular momentum ($\propto\sqrt{a_{\rm in}(1-e_{\rm in}^2)}$)
is roughly conserved during migration so the final
semi-major of the hot Jupiter in a circular orbit is 
$a_{{\rm in},f}\simeq2a_{\rm in}(1-e_{\rm in})$.}.

From panel a of Figure \ref{fig:emax_phi} we observe that
for $\mu=m_{\rm in}/m_{\rm out }\in[0.58,0.77]$ (equivalent
to $m_{\rm out}=[1.3,1.7]M_J$ as in the synthesis study) 
and  $e_{\rm in}=0.6$ the maximum eccentricities in the
range $0.95-0.985$ and the exact value increases
with the initial eccentricity of the outer planet $e_{\rm out}$.
Since the simulation starts with $e_{\rm out}$ taken from a 
Rayleigh distribution (Eq. [\ref{eq:sigma_e}]), then 
it is more likely for the inner planet
to reach lower maximum eccentricities and larger
pericenter distances in this range and, therefore, the
HJs would tend to have higher semi-major axes.
This result qualitatively explains why the semi-major axis 
distribution in the simulation does not peak at the smallest
allowed values.

From Figure \ref{fig:hj_sma}  we observe that our numerical study
mostly forms HJs with $a_{\rm in}<0.07$ AU, while 
 $\sim7\%$ and $\sim19\%$ of the observed HJs have
$a_{\rm in}>0.07$ AU in transit and RV surveys, respectively.
From the lower panel we observe that by restricting our
sample to HJs with $a<0.07$ AU, our population study describes 
the observed distribution fairly well ($p-$values $\gtrsim0.1$).

The observed population of HJs with $a>0.07$ AU can be explained
by CHEM by 
increasing the efficiency of tidal dissipation, which might
be achieved by either decreasing the planetary 
viscous time $t_{V,p}$ or considering an initially inflated planet as
in \citet{petro15}. 

\begin{figure}[h!]
   \centering
  \includegraphics[width=8.6cm]{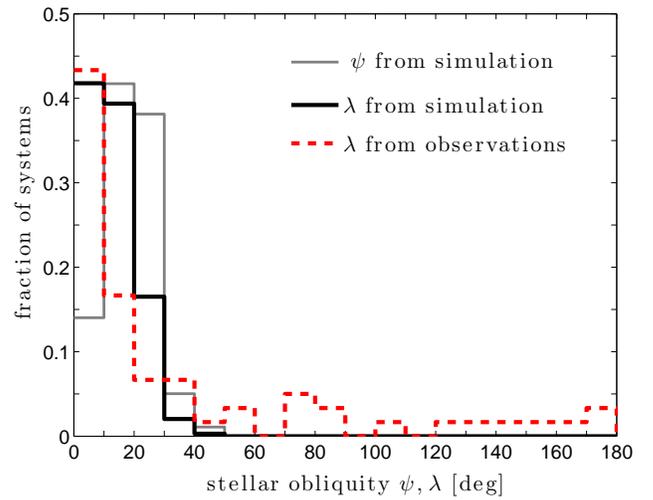} 
  \caption{Distribution of the stellar obliquity $\psi$ (gray
 solid line) and sky-projected stellar obliquity $\lambda$
  (black solid line) of the hot Jupiters formed
  our population synthesis study.  
  The sample of 60 hot Jupiters with $m\sin i>0.1M_J$ and
  projected obliquity measurements is shown in red dashed line.\\
  }
\label{fig:hj_psi}
\end{figure}  
\subsection{Obliquity distribution
of hot Jupiters}

As of September 2014, the observed sample of 
hot Jupiters\footnote{From The Exoplanet Orbit Database 
\citep{wright11} } (planets with $M \sin(i) > 0.1M_J$ and 
$a < 0.1$ AU) contains 60 planets
with projected stellar obliquity measurements
$\lambda$ with mean and median of
$\simeq38^\circ$ and $\simeq14^\circ$.

In Figure \ref{fig:hj_psi} we show the 
distribution of obliquities $\psi$ and projected 
obliquities $\lambda$ from our population synthesis study 
and compare this with the observed data.
From our simulations we measure
the angle between the spin axis of the host
star and the normal of the inner planetary orbit $\psi$
(often called the stellar obliquity 
angle or misalignment angle).
We then calculate $\lambda$, the sky-projected value of $\psi$,  
by taking $10^5$ random orbital configurations relative to a fixed 
observer for each system (see e.g., \citealt{FW09}).

We observe that the final distribution of $\psi$ is concentrated towards
$10^\circ-30^\circ$, while the HJ systems initially have zero 
obliquity and a low mutual inclination $i_{\rm tot}$ 
(mean and median of $\sim7^\circ$). 
Similar to our example in Figure \ref{fig:example_sec} the moderate 
excitation of $\psi$ comes from the excitation of  $i_{\rm tot}$ during the
high-eccentricity phases of the system's evolution.
Thus, the range of $\psi$ in HJ systems formed by CHEM depends
on range of the initial mutual inclination.
We checked this conclusion by considering initially
flatter systems (lower values of $i_{\rm tot}$), 
and indeed found that 
the distribution of $\psi$ shifts to lower values.

In Figure \ref{fig:hj_psi} we observe that our population synthesis study
of CHEM produces HJs with $\lambda<40^\circ$ and typically
($\sim80\%$) $\lambda<20^\circ$. 
This result compares favorably with the data because
most planets ($\sim60\%$) in the observations have 
$\lambda<20^\circ$. 
However, CHEM fails to explain the systems with
$\lambda>40^\circ$, which correspond to $\sim25\%$ of the
observed sample.

These systems with higher obliquities must be produced by 
another mechanism such as  
the  Kozai-Lidov mechanism in stellar binaries
(e.g., \citealt{WM03,FT07,naoz12,petro15}), planet-planet 
scattering (e.g., \citealt{NIB08,NI11,BN12}), or other secular 
interactions
between planets  (e.g., \citealt{naoz11,WL11}).
The higher obliquities can also be due to a primordial 
 misalignment of the proto-planetary 
disk relative to the host star's spin axis
(e.g., \citealt{BLP10,LFL11,B12,CB14,SB14}) or
a tilt of the outer layers of the host stars
\citep{rogers12,rogers13}.

We note that any primordial alignment of the stellar spin 
axis from the proto-planetary disk in which the proto-hot Jupiter 
is ultimately formed would be nearly preserved for the planetary 
orbit undergoing CHEM.
This property has been previously attributed to the hot Jupiters
formed through disk-driven migration since these planets remain
in the same plane as a the proto-planetary disk during migration.
Our simulations show that high-eccentricity
migration can also preserve the alignment between the 
stellar spin and planetary orbits.

\subsection{Migration timescale of hot Jupiters}
\label{sec:timescale}

From panel c in Figure \ref{fig:montecarlo} we observe 
that the migration timescale $t_{\rm mig}$ 
(or stopping time in the simulation)  of hot Jupiters in circular
orbits (red circles) increases monotonically with
the final semi-major axis $a_{\rm in}$.

We show that the empirical expression
$\log\left( t_{\rm mig}/1 {\rm Myr}\right)= 0.82\times a_{\rm in}/0.01~{\rm AU}-2$
(black dashed line) gives a good fit to the 
to the migration timescale as a function
of the final semi-major axis $a_{\rm in}$.
This expression is only valid for the parameters used in
our population synthesis study:
a Jupiter-like planet ($R_{\rm in}=1R_J$, $m_{\rm in}=1M_J$)
with a viscous time of the planet $t_{V,p}=0.1$ yr orbiting a
Sun-like star ($R_{\rm 1}=1R_\odot$, $m_{\rm 1}=1M_\odot$).
From \citet{petro15} (Equation 10 therein) we have that the
migration timescale depends on these parameters as
$t_{\rm mig}\propto t_{V,p}\left(m_{\rm in}/m_1\right)^2R_{\rm in}^{-8}$,
implying that the migration timescale
can be written as
\ba
\log\left( t_{\rm mig}/1 {\rm Myr}\right)&=& 0.82\times 
\frac{a_{\rm in}}{0.01~{\rm AU}}+
\log\left(\frac{t_{V,p}}{0.1~{\rm yr}}\right)\nonumber\\
&+&2\log\left(\frac{m_{\rm in}}{m_1}\frac{M_\odot}{M_J}\right)-
8\log\left(\frac{R_{\rm in}}{R_J}\right)-2.
\label{eq:tmig}\nonumber\\
\ea

This timescale can be used to compare this migration 
scenario with observations: 
a HJ with a given semi-major axis $a$ should be
older than $t_{\rm mig}$ from our fit.
For instance, \citet{quinn12} recently discovered two HJs
 in the 600 Myr Beehive cluster, with current semi-major
 axes 0.032 AU and 0.052 AU.
 Our empirical formula gives a minimum migration 
 timescale\footnote{ We use the fiducial parameters
 in Equation (\ref{eq:tmig}) since the HJs radii and masses
 have not been yet measured.} 
 of  $\sim4$ Myr and $\sim180$ Myr, respectively.
Thus, the migration timescales predicted by our population
synthesis study of CHEM are both within the age of the 
cluster.
 
Finally, we note that we stop the simulation when the HJ
reaches an eccentricity $e<0.01$, while the subsequent tidal
dissipation in the star can change the semi-major axis of the planet,
specially for short-period ($<3$ days) planets.
However, most stars hosting HJs have rotation periods longer
than $\sim 3$ days and, therefore, tides in the star are 
expected to shrink the semi-major axis of these 
short-period planets, making 
our constraint of the minimum timescale 
still valid.

\subsection{Outer planets in hot Jupiter systems}
\label{sec:outer}

The outer planets in our simulated 
hot Jupiter systems initially have moderately
high eccentricities (mean and median of 0.53 and 0.54), while
at the end of the simulations they 
have somewhat lower eccentricities
(mean and median of 0.32 and 0.33).
This reduction in the eccentricity of the outer planet is expected
because HJs are only formed when they lose almost all their angular 
momentum, which is mostly transferred to the orbit of the outer planet.

We expect that the outer planets with larger semi-major axes 
and larger masses
are less affected by this reduction in eccentricity 
because they have higher initial angular momentum
and, therefore, can retain larger final eccentricities. 
In particular, we observe a strong positive correlation 
(coefficient of $\simeq0.54$) between $e_{\rm out}$ and 
$a_{\rm out}$ in our simulations.

The outer planets in HJ systems have an initial mean
mutual inclination of $\simeq7.7^\circ$, which decreases 
slightly to $\simeq6^\circ$ once the HJ is formed.

We have restricted our population synthesis study to a limited
range in semi-major axes and masses of the outer body because
the parameter space is large and the initial conditions 
are fairly uncertain.
We can, however, place constraints on these parameters  
based on our analytical calculations 
in \S\S \ref{sec:circular} and  \ref{sec:ini_ecc},
 where we show in Equations (\ref{eq:mu_alpha}) and (\ref{eq:mu_alpha2})
that CHEM operates with the minimum eccentricity
of the outer planet for
\ba
m_{\rm out}a_{\rm out}^{1/2}=\tilde{C}\times m_{\rm in}a_{\rm in}^{1/2},
\label{eq:in_out}
\ea
where $\tilde C\simeq3.3$  and $\tilde C\simeq6.2$ for an initial inner planet
with zero eccentricity and with eccentricity equal to that of the outer
planet, respectively\footnote{Note that the other limit of an inner planet 
in a initially highly eccentric orbit ($e_{\rm in}\gtrsim0.8$) allows for a wider 
range of semi-major axis and mass ratios.}. 
The approximation that CHEM mostly operates with the minimum
eccentricity of the outer planet is justified if its distribution decreases
rapidly for $e_{\rm out}\gtrsim0.5$, as is observed
in the sample of giant planets at $a>1$ AU.

On the other hand, the dynamical stability of the system
requires that $a_{\rm out}/a_{\rm in}\gtrsim8$ and 
$a_{\rm out}/a_{\rm in}\gtrsim5$ for an initial inner planet 
with zero eccentricity and with eccentricity equal to 
that of the outer planet, respectively from these 
initial conditions, respectively. 
Note that by increasing $a_{\rm out}/a_{\rm in}$ CHEM becomes less
efficient since both the GR and the tidal quadrupole strongly limit
the eccentricity growth (see Eqs. [\ref{eq:gr_2}] and [\ref{eq:tide_2}])
and CHEM requires higher outer eccentricities to operate
(see Figure \ref{fig:e_max_mass}).
Thus, the most likely initial semi-major axis ratios are probably
close to $a_{\rm out}/a_{\rm in}\sim5-8$.

Roughly speaking, from the arguments above we conclude
that if the inner planet commenced CHEM at $a_{\rm in}\sim1$ AU, 
the most likely properties of outer planet are 
$a_{\rm out}\sim5-8$ AU and 
$m_{\rm out}/m_{\rm in}\sim1-3$ (from Eq. [\ref{eq:in_out}]).

In summary, the outer planets in HJ systems formed by CHEM
have moderate eccentricities ($e_{\rm out}\sim0.2-0.5$) 
and low mutual inclinations relative 
to the HJ's orbit. Their eccentricities are expected to be larger
for outer perturbers at wider separations or with higher masses.
Based on the minimum initial eccentricities required for
CHEM to operate we determine the most likely semi-major
axis and mass ratios to be $a_{\rm out}/a_{\rm in}\sim5-8$ and 
$m_{\rm out}/m_{\rm in}\sim1-3$,
respectively.

\section{Discussion}
 
\subsection{Comparison with previous work}

\subsubsection{Retrograde vs prograde HJs from 
CHEM}
 
We have shown that CHEM produces hot Jupiters in 
prograde and low obliquity orbits (assuming an initially 
zero misalignment of the 
planetary orbit relative to host star spin).
On the contrary, \citet{li14} concluded that CHEM is a mechanism
to produce counter-orbiting hot Jupiters 
(obliquities of $\sim180^\circ$).

We understand this difference from the necessary condition
to flip the orbit from prograde to retrograde, which is that eccentricity
forcing mechanism studied here can produce extremely high 
eccentricities: $1-e_{\rm in}\lesssim10^{-3}-10^{-4}$ \citep{li14}.
In order for this to happen the migrating planet has to be initially
placed at large enough semi-major axis to satisfy 
the following requirements:
\begin{itemize}
\item the planet does not get  tidally disrupted by reaching 
pericenter that are too close to the host star.
For instance, according to \citet{GRL11} a Jupiter-like planet
orbiting a sun-like star gets disrupted if
$a_{\rm in}(1-e_{\rm in})\lesssim0.013$ AU, which implies that
that the planet should start at $a_{\rm in}\gtrsim13$ AU to avoid
disruption when $1-e_{\rm in}\lesssim10^{-3}$.
\item extra precession forces (e.g., GR precession)
do not efficiently  limit the eccentricity growth. 
As discussed in \S\ref{sec:max_e_extra}, the maximum 
eccentricity depends on $a_{\rm in}$ as 
$1-e_{\rm in,max}\propto a_{\rm in}^{-2}$ 
($1-e_{\rm in,max}\propto a_{\rm in}^{-10/9}$) if the dominant 
precession source is GR (the tidal quadrupole).
Thus, all other things being equal, the maximum eccentricity 
can reach higher values for larger semi-major axes.
\end{itemize}

In this work, we have considered an initial semi-major
axes in $a_{\rm in}=1-5$ AU and the effects from GR precession,
 tides, and tidal disruptions.
Therefore, the maximum eccentricity is not high to allow 
for orbit flipping  (see maximum eccentricities 
in Figure \ref{fig:emax_phi}), although it does allow for moderate
excitation (up to $\sim20^\circ$) of the mutual inclination
between the orbits (see panel c in Figure \ref{fig:example_sec}). 

In the systematic study of coplanar flips by \citet{li14}, the authors 
ignore the effects from 
tides and tidal disruptions, and 
consider the test particle 
limit $\mu\ll1$ for which the extra precession forces like
GR become much less efficient than the planetary regime 
considered here ($\mu\sim1$)
at limiting the maximum eccentricity growth.
Recall from the arguments in \S\ref{sec:max_e_extra} 
that $1-e_{\rm in,max}\propto \mu^{2}$ if GR is the dominant 
precession force.
All these approximations allow for the inner eccentricity 
to reach extremely high values and flip. 
The authors do consider the effect of tides and tidal 
disruptions in one example of an orbit flip
(Figure 7 therein), but in this example the inner
planet is initially placed at large 
enough distances ($a_{\rm in}\sim40$ AU) that it can avoid
both being tidally disrupted and having the eccentricity growth
efficiently limited by extra precession forces.

In summary, CHEM generally produces hot Jupiters with low 
obliquities.
It might, however, produce highly mis-aligned hot Jupiters
provided that the migrating planet starts migration from a
large ($\gg1$ AU) semi-major axis. 

\subsubsection{Other secular high-eccentricity 
migration scenarios}

Various high-eccentricity migration mechanisms have been 
shown to produce hot Jupiters from gravitational interactions 
between planets like in CHEM. We briefly comment on the main
differences between these mechanisms and CHEM.

First, hot Jupiters can be formed by the chaotic 
secular interactions between two or more
planets in eccentric and/or mutually inclined orbits, proposed 
and termed secular chaos by \citet{WL11}. 
Here, the eccentricity excitation is chaotic and depends on the 
mutual inclination between planets since coplanar systems 
become much more regular. On the contrary, the eccentricity 
excitation from CHEM is regular (non-chaotic) and does 
not depend on the initial mutual inclination provided that it is 
not too high ($\gtrsim20^\circ$).
Both CHEM and secular chaos predict that hot Jupiters
should have distant planetary companions.
CHEM requires of only one companion, while secular chaos
does favor having two or more planetary companions
because the system has more degrees of freedom.

Second, hot Jupiters might be formed by the Kozai-Lidov (KL)
mechanism \citep{naoz11,naoz12,petro15}.
Unlike CHEM, the KL mechanism would require that the 
planetary orbits have initially high ($\gtrsim50^\circ$) 
mutual inclinations  (e.g., \citealt{teyss13}).
Also the eccentricity excitation in CHEM happens in
the octupole timescale that is longer 
by a factor of $\sim a_{\rm out}/a_{\rm in}$ than
the quadrupole timescale that governs 
the KL mechanism.
This slower eccentricity excitation allows  for extra 
precession  forces such as GR and tides
to limit the maximum eccentricity growth 
more efficiently, leading to the formation of
hot Jupiters with semi-major axes generally larger
than those expected from KL migration.

All the mechanisms above, including CHEM, 
require an initial configuration with well-spaced and 
eccentric or/and mutually inclined orbits either
of additional planets or stellar companions.
Given  that we do not know the initial 
states of planetary systems, it is difficult to assess which
mechanism is more likely to be prevalent.
One natural candidate to explain
the initial conditions required for these 
different high-eccentricity migration scenarios 
is planet-planet scattering starting from  
initially unstable planetary systems
(e.g., \citealt{JT08,CFMS2008}). 
We plan to address which set of initial 
conditions are more likely to emerge from scattering
in a future work (Petrovich \& Tremaine 2015, in prep.)

\subsection{Summary of 
predictions by CHEM}

We have shown that CHEM can produce hot Jupiters.
Whether CHEM produces most hot
Jupiters is a more difficult issue to address since we 
do not know the initial states of planetary
systems.
However, we can partly address this issue by 
comparing the predictions from CHEM with the
available (or upcoming) observations.

Coplanar High-eccentricity Migration predicts:
\begin{enumerate}
\item {\it a pile-up of hot Jupiters
at $a\sim0.04-0.05$ AU.}

This pile-up is a natural consequence from CHEM since
it  excites the eccentricity of the migrating planet very slowly
(slower than the Kozai-Lidov mechanism by a factor
$\sim a_{\rm out}/a_{\rm in}$) allowing for 
pericenter  precession forces due to general relativity and 
tides to efficiently limit the maximum 
eccentricity growth (see Figure \ref{fig:emax_phi}). 
This limit in the eccentricity translates into a minimum
pericenter distance and the formation of a hot Jupiter
with semi-major axis roughly twice this minimum distance, 
as discussed in \S\ref{sec:hj_sma}.

This predicted concentration of hot Jupiters with 
$a\sim0.04-0.05$ AU compares well with the observations 
of hot Jupiters detected in transit and RV surveys
(see Figure \ref{fig:hj_sma}).

\item {\it hot Jupiters with low stellar obliquities.}

The low stellar obliquities of HJ systems are a natural
consequence of
CHEM since the eccentricity of the migrating planet
can be excited to high values without exciting 
its inclination. 
This result shows that,  like disk-driven migration,
high-eccentricity migration can also preserve the alignment 
between the stellar spin and planetary orbits.
 
 Our population synthesis study shows that CHEM mostly 
 produces HJs with projected obliquities $\lesssim30^\circ$, and 
 almost $70\%$ of the current observations fall into this
range.
The remaining population of mis-aligned hot Jupiters might be 
explained by either another high-eccentricity migration channel
or a mechanism that tilts the star or the plane 
of the planetary system.

\item {\it a few percent occurrence rate of hot Jupiters per 
distant giant planet}. 

Our population synthesis study shows that $\sim3\%$ of the
systems produce a hot Jupiter. 
This number mostly depends on the initial eccentricities since
most HJs are formed starting from $e_{\rm in}>0.5$ and 
$e_{\rm out}\sim0.4-0.7$, a range containing 
$\sim15\%$  and  $\sim20\%$ of the known 
planets with $a>1$ AU, respectively. 
This fraction can increase by: 
\begin{itemize}
\item shifting the stability boundary for hierarchical triple
systems (Eq. [\ref{eq:stability}]) towards higher eccentricities.
Indeed, we repeated the population synthesis study using 
the less restrictive stability condition
$a_{\rm out}(1-e_{\rm out})>1.7a_{\rm in}(1+e_{\rm in})$
from \citet{EK95} and observed that the
occurrence rate increased from $3.1\%$ 
in our study using Equation (\ref{eq:stability})
to $5.2\%$;
\item starting with positively 
correlated inner and outer eccentricities, which might
be expected from an initial scattering phase.
\end{itemize}

If CHEM dominates the formation of HJs then
the ratio between the number of HJs and the number of gas 
giant planets should be $\sim3-5\%$.
This ratio is roughly consistent with the one 
derived from observations
of $\sim3-10\%$ since the occurrence rate of HJs is 
$\sim0.5-1.5\%$ \citep{gould06,mayor11} and that
of the gas giant planet at AU distances is $\sim15\%$
\citep{mayor11}.

\item {\it hot Jupiters have distant  massive 
companions in nearly coplanar and moderately eccentric
orbits}.

The most likely outer planets in HJ systems formed by CHEM
have moderate eccentricities ($e_{\rm out}\sim0.2-0.5$),
 low inclinations ($<10^\circ$) relative 
to the HJ's orbit, and masses $\sim1-3$ times
larger than that of the HJ (see \S\ref{sec:outer}).
Also, the most likely semi-major axis ratio before 
commencing migration is 
$a_{\rm out}/a_{\rm in}\sim5-8$ so assuming that
CHEM started at $a>1$ AU, we expect companions
at $a\gtrsim5$ AU.

Since the RV surveys have characterized giant planets with
full orbits up to $a\sim5$ AU, we expect that most of the 
companions predicted by CHEM generally appear as RV linear 
trends.
Recently, \citet{knutson14} estimated that $51\%\pm10\%$ 
of the HJs have a companion with $a_{\rm out}=1-20$ AU and 
masses of $m_{\rm out}=1-13 M_J$, while the masses of the
planetary companion tend to be comparable to or larger than
the transiting HJs.
This range of planetary masses and semi-major axes is
consistent with CHEM.

There are three systems with hot Jupiters and an
outer companion with eccentricity and
semi-major axis measurements\footnote{From www.exoplanets.org}:
\begin{itemize}
\item HD 217107 contains a HJ at 
0.075 AU with $e\simeq0.12$ and $m\sin i\simeq 1.41M_J$,
 and a companion at  6.07 AU with
$e\simeq0.38$ and $m\sin i\simeq 4.5M_J$ \citep{vogt05,feng15},
\item HD 187123  contains a HJ at 
0.042 AU in a circular orbit with $m\sin i\simeq 0.51M_J$
 and a companion at  $\simeq4.4$ AU with
$e\simeq0.28$ and $m\sin i\simeq 1.8M_J$
\citep{vogt05,wright09,feng15},
\item HAT-P-13  contains a HJ at 
0.043 AU in a circular orbit with $m\simeq0.85M_J$ 
and a companion at 
1.22 AU with $e\simeq0.66$ and $m\sin i\simeq 14M_J$
\citep{bakos09}. 
The hot Jupiter has a projected obliquity
of $\lambda=1.9^\circ\pm8.6^\circ$
\citep{winn10}.
\end{itemize}

We observe that the eccentricities and 
mass ratios (assuming nearly coplanar orbits) 
from HD 217107 and HD 187123 are 
roughly consistent with the most likely range predicted
from CHEM of $e\sim0.2-0.5$ and $\sim1-3$, respectively.
This result suggests that CHEM might have operated to form 
these close-in  planets. Migration in these systems 
should have commenced 
within $\sim1$ AU since the companions
are at $a\sim 5$ AU. 

On the contrary, HAT-P-13 has a mass ratio $>17$ 
and the perturber is at $\sim1$ AU, making CHEM 
an unlikely formation scenario. 
Moreover, given the high eccentricity of the outer planet
the stability boundary of hierarchical triple systems 
in Equation (\ref{eq:stability}) constrains the inner planet 
to $a\lesssim0.1$ AU and, therefore,  inconsistent 
with any high-eccentricity migration scenario.

\item {\it HJ formation timescales that increase
exponentially with semi-major axis.}

From our population synthesis study 
we find that the minimum timescale to form a hot Jupiter
depends exponentially with semi-major and
found an empirical fit given by
Equation (\ref{eq:tmig}).
As discussed in \S\ref{sec:timescale}, this minimum formation
timescale for the two hot Jupiters
in the Beehive cluster is consistent with 
its age 600 Myr \citep{quinn12}.
Future age constraints from hot Jupiter systems 
might prove useful to constrain CHEM.

More generally speaking, this minimum formation 
timescale from CHEM implies that
the occurrence rate of hot Jupiters 
should increase with stellar age and that the 
hot Jupiters with
larger semi-major axes should be restricted to older
systems. 
The former observation is consistent with the
difference between the
HJ abundances in Kepler and RV surveys
(e.g., \citealt{Dawson13}).

\item {\it a population of eccentric and low-obliquity
close-in planets}

Depending on the age of the planetary system
and the efficiency of tidal dissipation, CHEM is expected
to produce planets which have experienced significant
orbital migration, but have not had enough time to
become a HJ in a circular orbit
(see planets with final $e_{\rm in}>0.4$ and
$a_{\rm in}<0.3$ AU in Figure \ref{fig:montecarlo}).
These planets are eccentric and have relatively low stellar
obliquities ($\psi\lesssim20^\circ$).

There are three planetary systems with a 
 giant planet with $a<0.3$ AU, eccentricity 
of $e>0.4$, and with a measurement of 
its projected stellar obliquity $\lambda$:
\begin{itemize}
\item HD 17156  b is a planet with mass 
$\simeq3.3M_J$ at $a=0.16$ AU with eccentricity
$e\simeq0.68$, and projected obliquity 
$\lambda=10^\circ\pm5^\circ$
\citep{fischer07,narita09},
\item HAT-P-2 b is a planet with mass 
$8.9\pm0.4M_J$ at $a=0.068$ AU with eccentricity
$e\simeq0.52$, and projected obliquity 
$\lambda=10^\circ\pm5^\circ$
\citep{bakos07,albrecht12},
\item HAT-P-34 b is a planet with mass 
$3.3\pm0.2M_J$ at $a=0.068$ AU with eccentricity
$e\simeq0.44$, and projected obliquity 
$\lambda=0^\circ\pm14^\circ$
\citep{bakos12,albrecht12}.
\end{itemize}

We observe that all these three planets 
in close-in and eccentric orbits
have low projected obliquities ($\lambda\lesssim10^\circ$), 
suggesting that CHEM might have operated 
to form these systems.
This observation is particularly interesting because these planets
are hardly produced by other migration mechanism.
Other high-eccentricity migration mechanisms can produce 
high-eccentricity close-in planets 
similar to CHEM, but  these planets generally have higher 
 obliquities (e.g., \citealt{FT07,BN12}).
Similarly, disk-migration can naturally produce low-obliquity 
close-in planets, but neither disk-migration (e.g., \citealt{KN12}) nor
planet-planet scattering after migration to small 
orbital separations
are expected to excite high 
eccentricities \citep{johansen12,PTR14}.




\end{enumerate}

\section{Conclusions}

We study the secular gravitational interaction of
two planets in a hierarchical configuration with relatively 
low mutual inclinations and eccentric orbits, including the
effects from general relativity, tides, and stellar 
rotation.

We show that the eccentricity of the inner planet 
can be excited to very high values starting from:
an inner planet in a circular orbit and 
an outer planet with eccentricity of $\gtrsim0.67$ or
two eccentric orbits ($e\gtrsim0.5$). 
The excitation is most efficient 
(i.e., requires the smallest initial eccentricities) when 
the semi-major axis ratio $\alpha=a_{\rm in}/a_{\rm out}$ and 
mass ratio $\mu=m_{\rm in}/m_{\rm out}$ are in the following 
range $\mu(\alpha/0.1)^{1/2}\sim0.5-0.8$.

We show that this mechanism, which we term 
Coplanar High-eccentricity Migration (CHEM) 
can preserve the alignment between the stellar 
spin and the planetary orbits, 
generally forming hot Jupiters with low stellar obliquities.
Based on a population synthesis study we show
the hot Jupiters produced by CHEM can well-reproduce
the observed semi-major axis distribution of hot Jupiters
and  can account for their observed occurrence rates.

We predict that the hot Jupiters formed by CHEM
should have distant ($\gtrsim5$ AU) 
planetary companions in low mutual inclination and 
moderately eccentric ($e\sim0.2-0.5$) orbits
and with most likely masses 
$\sim1-3$ times larger than that of the HJ.\\

\acknowledgements 
 I acknowledge support from the 
CONICYT Bicentennial  Becas Chile fellowship. 
I am indebted to Scott Tremaine who has critically
and patiently read and commented on various versions 
of this paper.
I am also grateful to  Renu Malhotra, Smadar Naoz, Gongjie Li,
and Amaury Triaud for enlightening
discussions and comments, and
the anonymous referee for a very useful report.
All simulations were carried out using computers 
supported by the Princeton Institute of Computational 
Science and Engineering.


\end{document}